\documentclass[prl,lengthcheck,twocolumn,superscriptaddress]{revtex4-2}
\usepackage[utf8]{inputenc}
\usepackage{graphicx}
\usepackage{subfig}
\usepackage{amsmath, amsthm, amssymb, amsfonts, float, booktabs}
\usepackage{mathtools}
\usepackage{verbatim}
\usepackage{dcolumn}
\usepackage{bm}
\usepackage{epsf}
\usepackage{xcolor}
\usepackage{physics}
\usepackage{dsfont}
\usepackage{tikz}
\usepackage{todonotes}
\usepackage{multirow}
\usepackage{siunitx}
\usepackage{caption}
\usepackage{ragged2e}
\usepackage[percent]{overpic}
\usepackage{xr-hyper}

\usepackage[colorlinks=true,citecolor=blue,linkcolor=blue,urlcolor=blue]{hyperref}

\usepackage{cleveref}

\usepackage[font=small,labelfont=bf]{caption}

\DeclareCaptionJustification{justified}{\justifying}
\usetikzlibrary{arrows.meta,positioning,calc,decorations.pathmorphing}

\begin{document}

\title{Chirality routing non-polaritonic vacuum correlations in Landau polaritons}

\author{Ayoub El-Amrani}
\email{ayoub.el-amrani@um6p.ma}
\affiliation{College of Physical Sciences and Engineering, University Mohammed VI Polytechnic, Ben Guerir, 43150, Morocco.}

\author{Zakaria Mzaouali}
\affiliation{J\"ulich Supercomputing Centre, Institute for Advanced Simulation, Forschungszentrum J\"ulich, Wilhelm-Johnen-Straße, J\"ulich, 52428, Germany.}
\affiliation{Institut f\"ur Theoretische Physik, Universit\"at T\"ubingen, Auf der Morgenstelle 14, 72076 T\"ubingen, Germany.}

\author{Houssam Sabri}
\affiliation{College of Physical Sciences and Engineering, University Mohammed VI Polytechnic, Ben Guerir, 43150, Morocco.}
\author{Herschel Rabitz}
\affiliation{Department of Chemistry, Princeton University, Princeton, New Jersey 08544, USA.}
\author{Abdelouahed El Fatimy}
\affiliation{College of Physical Sciences and Engineering, University Mohammed VI Polytechnic, Ben Guerir, 43150, Morocco.}

\author{Dukhyung Lee}
\email{dukhyung.lee@um6p.ma}
\affiliation{College of Physical Sciences and Engineering, University Mohammed VI Polytechnic, Ben Guerir, 43150, Morocco.}

\begin{abstract}
Ultrastrong coupling between matter and cavity vacuum fields can turn the electromagnetic vacuum into a structured quantum environment, thereby opening passive routes for modifying and manipulating material properties. Recent work has identified light--matter entanglement as an important ingredient in these property changes, which raises the question of where the relevant vacuum correlations actually reside. Landau polaritons provide chiral ultrastrong coupling systems in which one circular cavity polarization forms the bright polariton branches. Here, using a quantum information approach, we show that an exact chiral charge in a multimode Hopfield model routes the dominant anomalous correlations, squeezing, and cavity–matter entanglement into the opposite polarization. We find that, using parameters extracted from a multimode Landau polariton system, this hidden sector correlates the cyclotron resonance with finite momentum magnetoplasmons through Gaussian discord, while pairwise matter–matter entanglement remains absent. We further predict a polarization anisotropy of dressed vacuum electric field fluctuations as a signature of this chiral routing. These results identify chirality as a symmetry principle for organizing ultrastrong coupling vacua and show that quantum information tools provide a powerful framework for revealing the salient properties of Landau polaritons.
\end{abstract}


\maketitle


\section{Introduction}

Ultrastrong light--matter coupling (USC) profoundly reshapes the quantum vacuum of cavity systems. When the interaction energy becomes comparable to the bare frequencies of the uncoupled subsystems, the rotating wave approximation breaks down, and both counter-rotating and diamagnetic terms become essential ingredients of the theory~\cite{FornDiazRMP2019,KockumNatRevPhys2019,SchaferACSP2020}. In this regime, photonic and material excitations hybridize non-perturbatively into polaritonic normal modes~\cite{Hopfield1958}, while the ground state is converted from an empty vacuum into a dressed Gaussian state hosting squeezed fluctuations and anomalous correlations~\cite{CiutiPRB2005,FornDiazRMP2019,DeLiberato2017Virtual}. Although these vacuum fluctuations do not appear as real photon populations in equilibrium~\cite{CiutiPRB2005}, they lead to solid physical impacts: response functions, spectral properties, and other measurable observables can be modified without requiring the extraction of real photons~\cite{PhysRevA.81.042311,DeLiberato2017Virtual,cxvs-5pb1,Baydin2025Perspective}. Direct access to such vacuum excitations can be implemented through dedicated probing protocols, including non-adiabatic modulation, spontaneous conversion schemes, electroluminescence proposals, and coherent transfer or open system detection strategies~\cite{DeLiberatoPRL2007,PhysRevLett.110.243601,Cirio2016,PhysRevLett.122.190403,Falci2019Ultrastrong,Ridolfo2021Probing}.

In solid state platforms, the physical consequences of the vacuum become particularly evident. In Landau polariton systems, cavity vacuum fields have been shown to modify the transport and collective properties of a two-dimensional electron gas (2DEG) in equilibrium~\cite{PhysRevB.98.205301,ParaviciniBagliani2019,Appugliese2022}. More directly, vacuum field correlations can now be accessed through field correlation measurements without relying on photon emission or population extraction~\cite{BeneaChelmusNature2019}. These developments establish USC ground states as structured quantum environments with observable consequences, motivating a question: how are the quantum correlations and field fluctuations of the dressed vacuum distributed among the underlying light and matter modes?

Landau polaritons provide a particularly rich setting for this question. A 2DEG subjected to a perpendicular magnetic field (c.f. Fig.~\ref{fig:symmetry}b) supports cyclotron and magnetoplasmon excitations that can couple to terahertz cavity modes, reaching the ultrastrong and even deep strong coupling regimes~\cite{ScalariScience2012,ZhangNatPhys2016,KellerPRB2020,MornhinwegNatCommun2024}. Because several collective electronic modes can couple to the same cavity resonance, the platform naturally realizes a multimode light--matter problem beyond the minimal case of a single light-matter pair. Recent works have experimentally demonstrated the multimode polaritonic spectrum, whose mode structure is accurately described by Hopfield-type models~\cite{MornhinwegNatCommun2024,EndoKimLiangLeeKimCovarrubiasMoralesSeoManfraLeeBambaKono+2025+4647+4654}. This raises a question that is complementary to spectroscopy: is cavity mediation in the ultrastrong coupling vacuum exhausted by the hybridization visible in the polariton spectrum, or does it also redistribute quantum correlations between local and nonlocal electronic modes?

More broadly, chiral light--matter interactions have become an increasingly active frontier across nanophotonics, condensed matter cavity QED, and quantum technologies, driven by the possibility of engineering chiral electromagnetic environments that control symmetry breaking, light, charge, and spin~\cite{VanOrman2025chiral,Tay2025Chiral,PhysRevB.109.L161302,1tlw-g26r}. A key feature of the Landau polariton platform is its intrinsic chirality. In a circular polarization basis, only one cavity polarization mode couples resonantly to matter, whereas the complementary polarization enters only through counter-rotating processes~\cite{Li2018,EndoKimLiangLeeKimCovarrubiasMoralesSeoManfraLeeBambaKono+2025+4647+4654}. In spectroscopic language, one sector is active, hosting the familiar avoided crossing polariton branches, while the complementary inactive sector does not generate the conventional polariton anticrossings, instead appearing through the vacuum Bloch--Siegert shift~\cite{Li2018}. What has remained unclear, however, is whether the same active/inactive separation also governs the equilibrium quantum vacuum, or whether it is only a feature of the spectroscopic branches.

Here we show that this polarization asymmetry originates from the structure of the Hopfield Hamiltonian. We identify a conserved chiral \(U(1)\) charge that imposes strict selection rules on anomalous correlators and therefore fixes the allowed symmetry channels through which ground state squeezing and entanglement can be distributed. The symmetry suppresses anomalous channels involving the spectroscopically active polarization while leaving the complementary counter-rotating channel symmetry allowed, as summarized in Fig.~\ref{fig:symmetry}a. The resulting vacuum structure is therefore \textit{non-polaritonic} in an operational sense: its dominant correlations are organized not by the active polarization that forms the observed polariton branches, but by the orthogonal counter-rotating sector of the same Hopfield Hamiltonian. We then investigate how this symmetry is realized in an experimentally verified model system of a multimode Landau polariton Hamiltonian using continuous variable quantum information diagnostics~\cite{WeedbrookRMP2012,ferraro2005gaussianstatescontinuousvariable,BraunsteinVanLoockRMP2005,EisertPlenio2003,BraunsteinKimblePRL1998,LloydBraunsteinPRL1999,MenicucciPRL2006}. This approach have played an important role in revealing correlation structures that are not directly visible in conventional observables~\cite{Latorre_2009,Osterloh2002Scaling,OsborneNielsen2002,Vidal2003Critical,Amico2008Review,Mzaouali2019XXChain,Mzaouali2019PhaseSpaceSpinChains,Hawary2024PhaseSpace,Abaach2023FermiHubbard,mzaoualipre}. Here we use the same philosophy in a bosonic light--matter vacuum to show how the dressed ground state is correlated in the underlying cavity and collective electronic degrees of freedom. Since the Hamiltonian is quadratic, this information is fully encoded in the Gaussian covariance matrix, allowing vacuum fluctuations, logarithmic negativity, Gaussian discord, and graph state nullifiers to identify the entangling channels, the nonclassical matter correlations, and the organized multimode squeezing of the ultrastrong coupling vacuum~\cite{MenicucciPRL2006,ZhangBraunstein2006,MenicucciPRA2011}. In this sense, the quantum information measures act as a probe of the symmetry organized vacuum structure of Landau polaritons. Recent driven dissipative work also identified light--matter entanglement and quadrature squeezing as characteristic observables of Landau polariton dynamics~\cite{Mivehvar2026PRL}, further supporting the use of quantum information diagnostics in this platform.

Using this approach, we find that the polarization dominating the polariton spectrum remains essentially separable from the electronic modes in equilibrium, whereas the orthogonal polarization acts as the main mediator of vacuum squeezing and quantum correlations. In this sense, the dominant vacuum structure is non-polaritonic: it is organized by the polarization sector that does not form the bright polariton branches. We further show that the cavity-mediated coupling between local and nonlocal matter modes appears in the equilibrium vacuum predominantly as Gaussian discord without detectable pairwise matter--matter entanglement. In other words, the cavity mediates genuinely nonclassical, yet non-entangling, correlations inside the matter sector: quantum correlations that cannot be reduced to locally accessible classical information, even though they do not take the form of pairwise entanglement. This redistribution is routed predominantly through the complementary counter-rotating polarization and, for the experimentally extracted parameter set, gives rise to a nearly pure, genuinely multipartite correlated subsystem consistent with a star-like Gaussian graph state interpretation (Fig.~\ref{fig:symmetry}c).

Finally, we translate the chiral routing into a field level signature. Because the two circular cavity modes are dressed unequally, the vacuum electric field fluctuations become anisotropic in polarization space. This polarization-resolved fluctuation anisotropy provides an observable fingerprint of the hidden non-polaritonic vacuum structure: the symmetry-selected redistribution of vacuum correlations enhances or suppresses field noise depending on the polarization.

Our findings elevate the polarization asymmetry of Landau polaritons from a spectroscopic observation to a symmetry principle governing the ultrastrong coupling vacuum. While the underlying polarization asymmetric Hamiltonian can be inferred from the polariton spectrum, the vacuum correlations themselves remain hidden from conventional spectroscopy. The chiral selection rule and quantum information diagnostics reveal where the equilibrium vacuum correlations actually reside, information that is not directly readable from the spectroscopic dispersion alone. This hidden organization may be probed through vacuum field correlation measurements and may manifest in polarization sensitive response and magnetotransport in cavity coupled electron systems~\cite{BeneaChelmusNature2019,Li2018,PhysRevB.98.205301,ParaviciniBagliani2019,Appugliese2022,Enkner2025}.

\section{Results}

\paragraph*{\textbf{Chiral symmetry selection rules for the vacuum correlation channel.}}

\begin{figure*}[t]
\centering

\begin{minipage}[c]{0.7\textwidth}
\centering
\begin{tikzpicture}
    \node[inner sep=0] (A)
    {\includegraphics[
        width=\linewidth,
        trim={0mm 0mm 0mm 0mm},
        clip
    ]{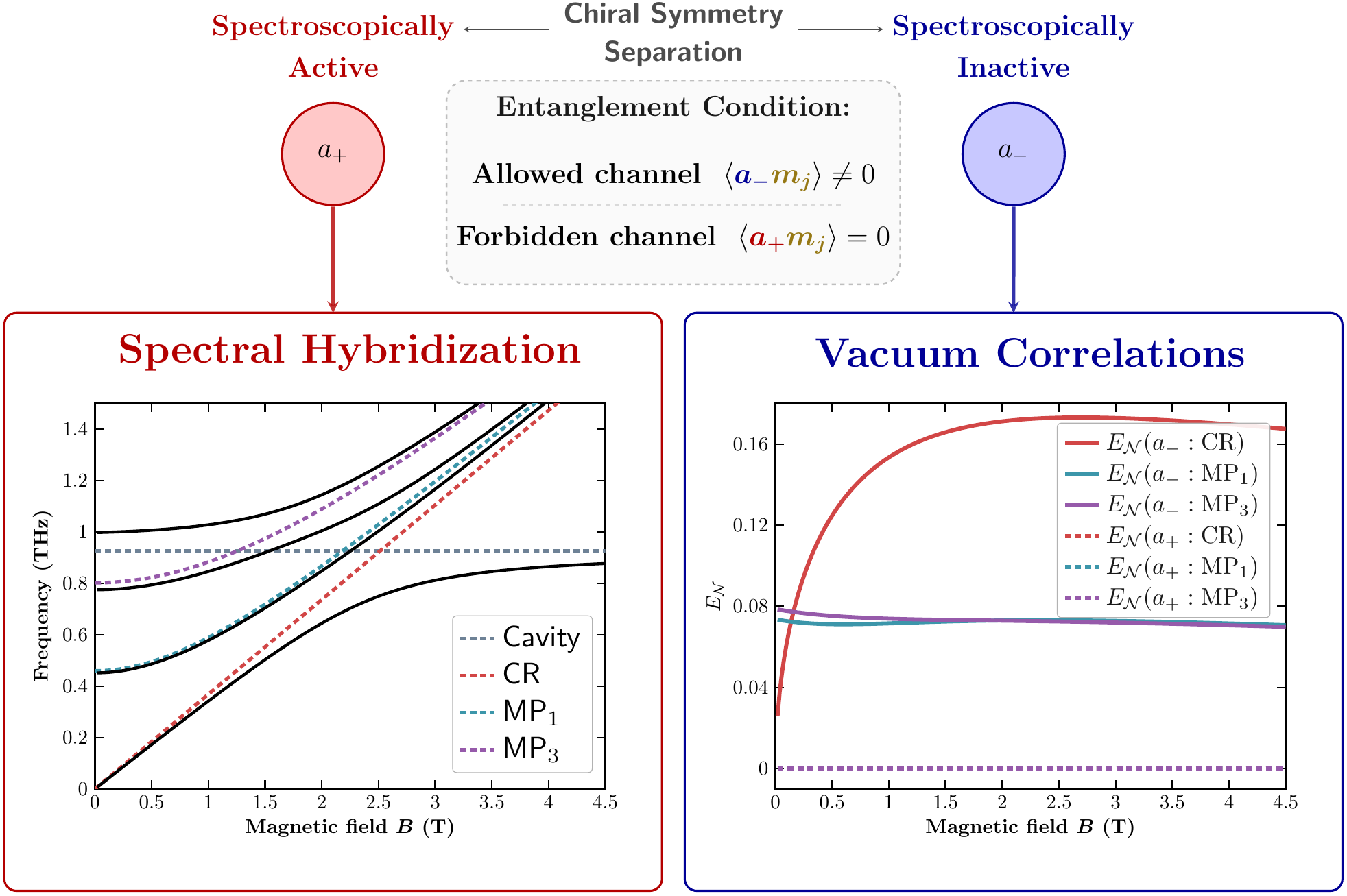}};
    \node[
        anchor=north west,
        font=\bfseries\fontsize{15}{15}\selectfont
    ] at ([xshift=0mm,yshift=-2mm]A.north west) {(a)};
\end{tikzpicture}
\end{minipage}
\hfill
\begin{minipage}[c]{0.292\textwidth}
\centering

\begin{tikzpicture}
    \node[inner sep=0] (B)
    {\includegraphics[
        width=\linewidth,
        trim={0mm 0mm 0mm 0mm},
        clip
    ]{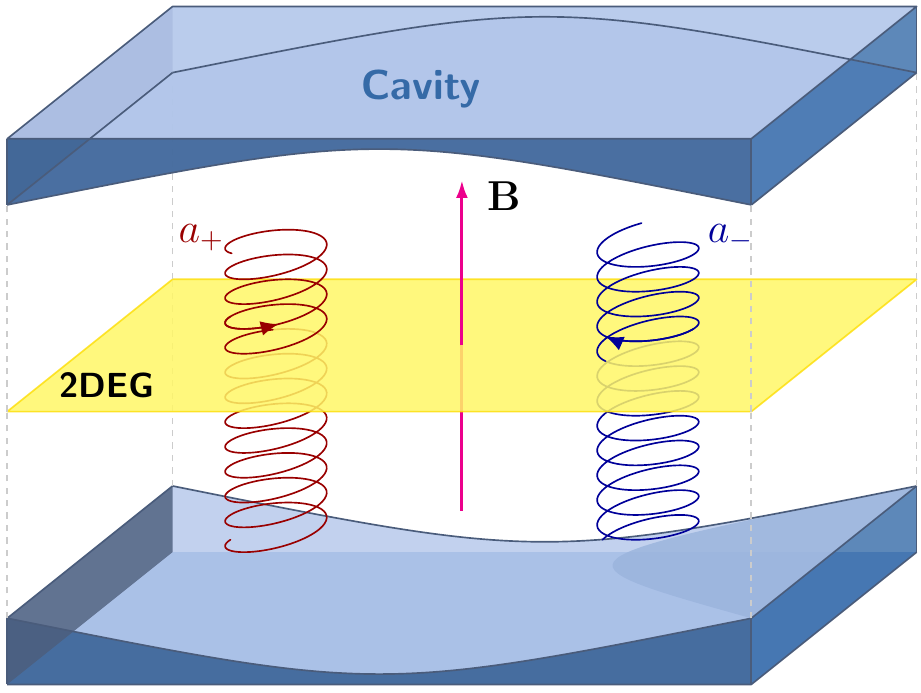}};
    \node[
        anchor=north west,
        font=\bfseries\fontsize{15}{15}\selectfont
    ] at ([xshift=-3mm,yshift=10mm]B.north west) {(b)};
\end{tikzpicture}

\vspace{2mm}

\begin{tikzpicture}
    \node[inner sep=0] (C)
    {\includegraphics[
        width=0.92\linewidth,
        trim={0mm 0mm 0mm 0mm},
        clip
    ]{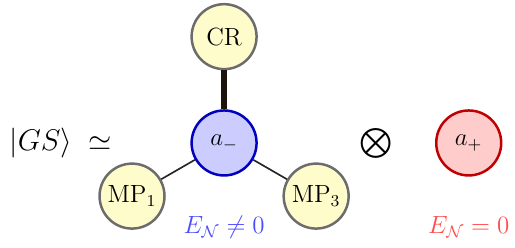}};
    \node[
        anchor=north west,
        font=\bfseries\fontsize{15}{15}\selectfont
    ] at ([xshift=-3mm,yshift=6mm]C.north west) {(c)};
\end{tikzpicture}

\end{minipage}

\caption{
\textbf{Chiral symmetry separating the bright polariton sector from vacuum correlation channels.}
\textbf{(a)} Structure of the multimode Landau polariton Hamiltonian, Eq.~\eqref{eq:Hschematic_results}. The circular polarization \(a_+\) is spectroscopically active and organizes the bright polariton hybridization, while the orthogonal polarization \(a_-\) hosts the anomalous vacuum correlation channels allowed by the symmetry. The chiral symmetry forbids anomalous \(a_+\)-matter correlations, \(\langle a_+ m_j\rangle=0\), while allowing the corresponding \(a_-\)-matter channels, \(\langle a_- m_j\rangle\neq0\). The spectral hybridization and the vacuum correlations quantified by logarithmic negativity \(E_{\mathcal N}\), Eq.~29 of Supplementary Note~5, were calculated based on the model system suggested in Fig.~\ref{fig:platform_spectral_benchmark}a.
\textbf{(b)} Schematic diagram of a Landau polariton system coupled with an optical cavity: a two-dimensional electron gas (2DEG) embedded in a terahertz cavity under a perpendicular magnetic field, supporting cyclotron and finite momentum magnetoplasmon excitations.
\textbf{(c)} Ground state correlation structure revealed in this work. The correlated sector is centered on \(a_-\), with the strongest graph-like connection to the cyclotron mode, whereas \(a_+\) remains nearly separable.
}
\label{fig:symmetry}
\end{figure*}
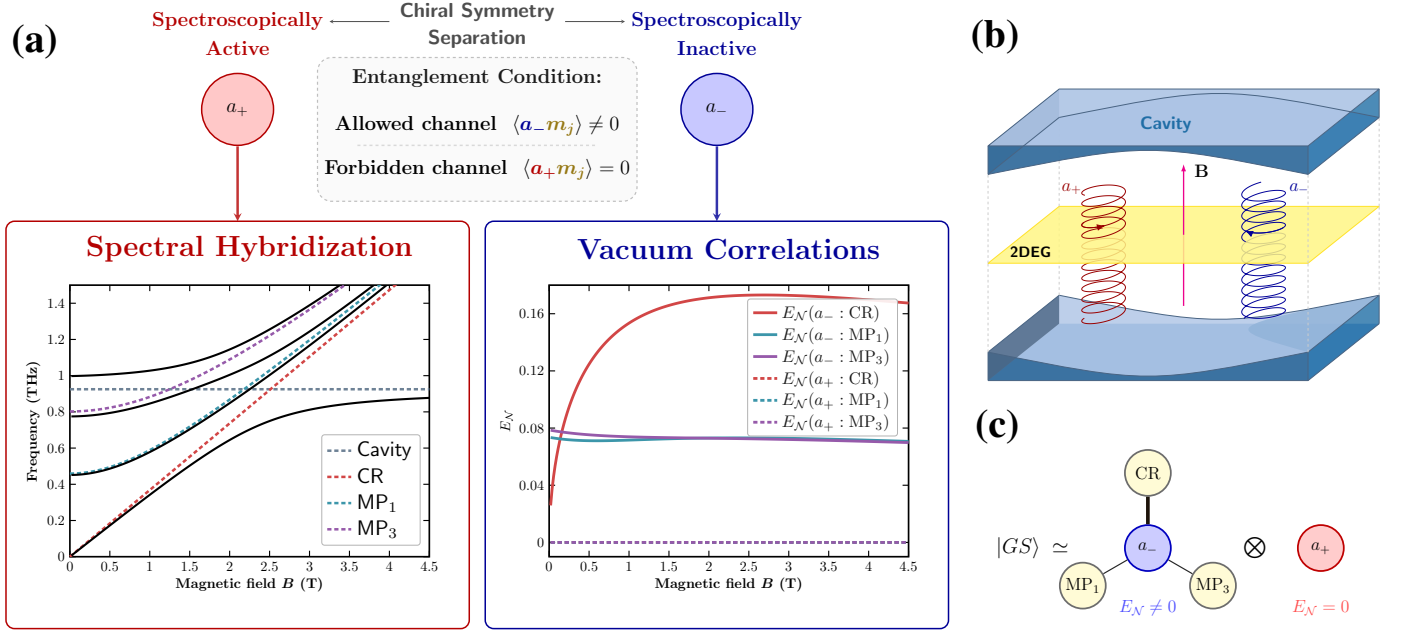

The central analytical result of this work is that the polarization-resolved Hopfield Hamiltonian possesses an exact chiral symmetry that already distinguishes the sector hosting the polariton branches from the sector carrying anomalous vacuum correlations. Related chiral light--matter physics beyond the rotating wave approximation has been investigated previously in a two-mode chiral Rabi model, where a copolarized mode couples through rotating wave terms while a counter-polarized mode is activated through counter-rotating processes under a conserved \(U(1)\) quantity associated with angular momentum~\cite{PhysRevLett.123.133603}. Much more abundant physics exists in Landau polariton systems: here the same chiral logic is realized in a solid state cavity QED platform described by a multimode Hopfield Hamiltonian, where the cavity couples not to a single two-level system but to collective bosonic excitations of a 2DEG, including both local and nonlocal modes, with the ability to continuously tune the interaction by a magnetic field. This makes it possible to address not only the chiral spectrum, but also how the symmetry reorganizes the equilibrium vacuum, rerouting squeezing and quantum correlations across the multimode matter sector. At the schematic level, the Hamiltonian can be written, in units of $\hbar=1$, as
\begin{equation}
\begin{split}
& H = \omega_0 \left(a_+^\dagger a_+ + a_-^\dagger a_-\right)
+ \sum_j \omega_j m_j^\dagger m_j \\
& + i \sum_j g_j
\Big[m_j^\dagger (a_+ + a_-^\dagger)
- m_j (a_- + a_+^\dagger)\Big] \\
& + D \,(a_- + a_+^\dagger)(a_+ + a_-^\dagger),
\end{split}
\label{eq:Hschematic_results}
\end{equation}
where $m_j$ denotes the collective matter modes.

A convenient way to expose the physical content of the Hamiltonian, Eq.~\eqref{eq:Hschematic_results}, is to inspect the operator sectors closed under the Heisenberg equations. The dynamics separates into
\begin{align}
\bm{\eta}_- = (a_+,m_1,m_2,\dots,a_-^\dagger)^T, \\
\qquad
\bm{\eta}_+ = (a_+^\dagger,m_1^\dagger,m_2^\dagger,\dots,a_-)^T,
\label{eq:eta_pm_results}
\end{align}
where the subscripts now follow the chiral charge of the sector. The active spectroscopic sector is therefore $\bm{\eta}_-$: it contains the cavity annihilation operator $a_+$ together with the matter annihilation operators $m_j$, i.e. the operators that can hybridize in the usual co-rotating, number conserving sense to form polaritonic normal modes. This is why the avoided crossing branches identified as polaritons live in the $a_+$ channel. By contrast, the inactive polarization enters the same sector only as the creation operator $a_-^\dagger$. It therefore does not participate in the usual spectroscopic hybridization, but instead appears through the antiresonant channel.

This sector decomposition is the dynamical signature of the conserved chiral charge
\begin{equation}
Q = a_+^\dagger a_+ + \sum_j m_j^\dagger m_j - a_-^\dagger a_- ,
\qquad
[H,Q]=0.
\label{eq:Q_results}
\end{equation}
Under the corresponding unitary transformation $U(\theta)=e^{-i\theta Q}$, the active polarization $a_+$ and the matter modes carry charge $-1$, while the inactive polarization $a_-$ carries charge $+1$. As a consequence, any anomalous operator with nonzero $Q$ charge has vanishing expectation value in a nondegenerate ground state. In particular,
\begin{equation}
\langle a_+ a_+ \rangle = 0,
\qquad
\langle a_+ m_j \rangle = 0,
\label{eq:selectionrule_results}
\end{equation}
for all matter modes $m_j$. The detailed derivation is given in Methods. A direct covariance matrix verification of the chiral selection rule is given in Supplementary Note 4.

Our key prediction is represented in Eq.~\eqref{eq:selectionrule_results}, because the Hopfield Hamiltonian is quadratic in bosonic operators, its equilibrium ground state is Gaussian and is therefore fully characterized by second moments, including the anomalous pair correlators that encode vacuum squeezing and correlation structure~\cite{WeedbrookRMP2012}. In continuous variable Gaussian systems, squeezing is the nonclassical resource underlying Gaussian entanglement~\cite{PhysRevA.64.063811,PhysRevA.65.032323,PhysRevA.66.024303}. In the present quadratic ultrastrong coupling Hamiltonian, this squeezing originates from the anomalous correlations generated by the counter rotating sector~\cite{CiutiPRB2005}. The active polarization may therefore organize the polariton spectrum, but the vacuum squeezing and entangling channels are symmetry forbidden in that sector and remain allowed only through the inactive polarization $a_-$. The chiral symmetry therefore implies that the dominant ground state correlations reside not in the active spectroscopic channel, but in the inactive polarization sector.

\paragraph*{\textbf{The multimode model with a prior experimental demonstration.}}

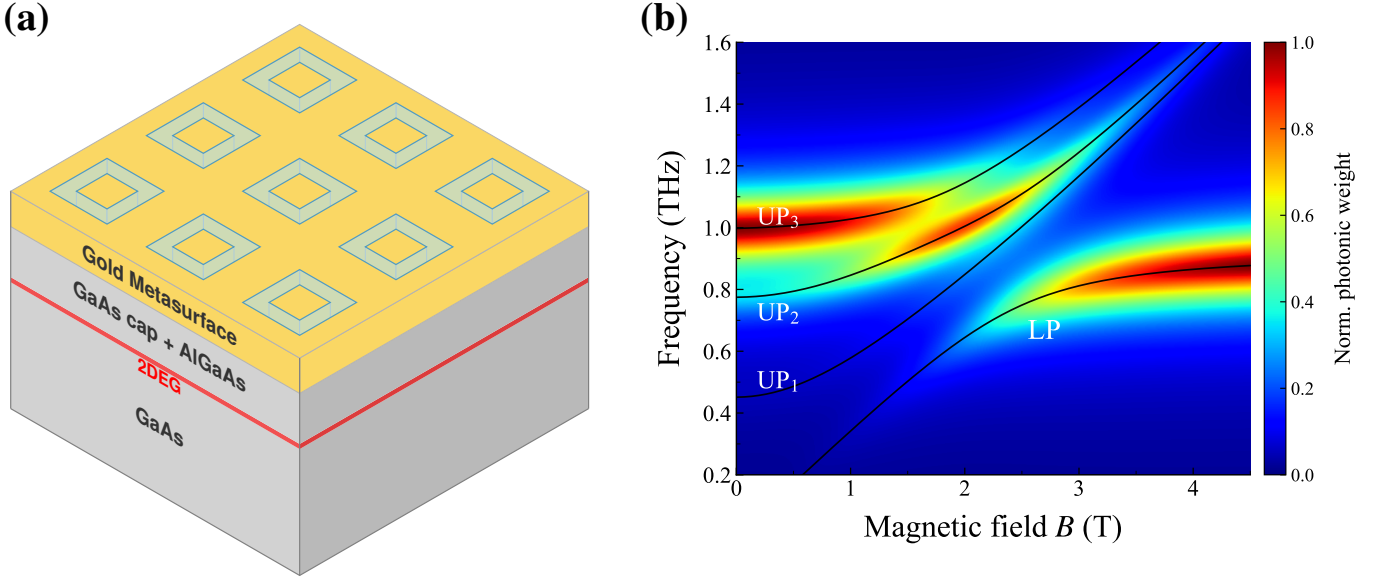
\begin{figure*}[t]
\centering

\begin{minipage}[t]{0.43\textwidth}
\centering
\vspace{0pt}

\begin{tikzpicture}
\node[inner sep=0] (A) {%
\resizebox{\linewidth}{!}{%
\begin{tikzpicture}[
    x={(-0.866cm, -0.5cm)},
    y={(0.866cm, -0.5cm)},
    z={(0cm, 1cm)},
    scale=1.2
]

\definecolor{gaastop}{RGB}{235, 235, 235}
\definecolor{gaasleft}{RGB}{210, 210, 210}
\definecolor{gaasright}{RGB}{185, 185, 185}
\definecolor{twodegleft}{RGB}{240, 80, 80}
\definecolor{twodegright}{RGB}{220, 60, 60}
\definecolor{goldtop}{RGB}{250, 215, 100}
\definecolor{goldleft}{RGB}{250, 215, 100}
\definecolor{goldright}{RGB}{250, 215, 100}
\definecolor{goldinnerX}{RGB}{250, 215, 100}
\definecolor{goldinnerY}{RGB}{250, 215, 100}

\def\L{6.6}
\def\gaasBot{0}
\def\gaasTop{2.5}
\def\twodegTop{2.6}
\def\capTop{3.6}
\def\goldTop{4.3}

\fill[gaasright] (0, \L, \gaasTop) -- (\L, \L, \gaasTop) -- (\L, \L, \gaasBot) -- (0, \L, \gaasBot) -- cycle;
\fill[gaasleft] (\L, 0, \gaasTop) -- (\L, \L, \gaasTop) -- (\L, \L, \gaasBot) -- (\L, 0, \gaasBot) -- cycle;

\fill[twodegright] (0, \L, \twodegTop) -- (\L, \L, \twodegTop) -- (\L, \L, \gaasTop) -- (0, \L, \gaasTop) -- cycle;
\fill[twodegleft] (\L, 0, \twodegTop) -- (\L, \L, \twodegTop) -- (\L, \L, \gaasTop) -- (\L, 0, \gaasTop) -- cycle;

\fill[gaasright] (0, \L, \capTop) -- (\L, \L, \capTop) -- (\L, \L, \twodegTop) -- (0, \L, \twodegTop) -- cycle;
\fill[gaasleft] (\L, 0, \capTop) -- (\L, \L, \capTop) -- (\L, \L, \twodegTop) -- (\L, 0, \twodegTop) -- cycle;
\fill[gaastop] (0,0,\capTop) -- (\L,0,\capTop) -- (\L,\L,\capTop) -- (0,\L,\capTop) -- cycle;

\foreach \cx in {1.1, 3.3, 5.5} {
    \foreach \cy in {1.1, 3.3, 5.5} {
        \pgfmathsetmacro{\xoneO}{\cx-0.65}
        \pgfmathsetmacro{\xtwoO}{\cx+0.65}
        \pgfmathsetmacro{\yoneO}{\cy-0.65}
        \pgfmathsetmacro{\ytwoO}{\cy+0.65}
        \pgfmathsetmacro{\xoneI}{\cx-0.35}
        \pgfmathsetmacro{\xtwoI}{\cx+0.35}
        \pgfmathsetmacro{\yoneI}{\cy-0.35}
        \pgfmathsetmacro{\ytwoI}{\cy+0.35}

        \fill[goldinnerX, draw=black!30, thin] (\xoneO, \yoneO, \goldTop) -- (\xoneO, \ytwoO, \goldTop) -- (\xoneO, \ytwoO, \capTop) -- (\xoneO, \yoneO, \capTop) -- cycle;
        \fill[goldinnerY, draw=black!30, thin] (\xoneO, \yoneO, \goldTop) -- (\xtwoO, \yoneO, \goldTop) -- (\xtwoO, \yoneO, \capTop) -- (\xoneO, \yoneO, \capTop) -- cycle;
        \fill[goldinnerX, draw=black!30, thin] (\xtwoI, \yoneI, \goldTop) -- (\xtwoI, \ytwoI, \goldTop) -- (\xtwoI, \ytwoI, \capTop) -- (\xtwoI, \yoneI, \capTop) -- cycle;
        \fill[goldinnerY, draw=black!30, thin] (\xoneI, \ytwoI, \goldTop) -- (\xtwoI, \ytwoI, \goldTop) -- (\xtwoI, \ytwoI, \capTop) -- (\xoneI, \ytwoI, \capTop) -- cycle;
    }
}

\fill[goldright, draw=black!30, thin] (0, \L, \goldTop) -- (\L, \L, \goldTop) -- (\L, \L, \capTop) -- (0, \L, \capTop) -- cycle;
\fill[goldleft, draw=black!30, thin] (\L, 0, \goldTop) -- (\L, \L, \goldTop) -- (\L, \L, \capTop) -- (\L, 0, \capTop) -- cycle;

\def\holepaths{}
\foreach \cx in {1.1, 3.3, 5.5} {
    \foreach \cy in {1.1, 3.3, 5.5} {
        \pgfmathsetmacro{\xoneO}{\cx-0.65}
        \pgfmathsetmacro{\xtwoO}{\cx+0.65}
        \pgfmathsetmacro{\yoneO}{\cy-0.65}
        \pgfmathsetmacro{\ytwoO}{\cy+0.65}
        \pgfmathsetmacro{\xoneI}{\cx-0.35}
        \pgfmathsetmacro{\xtwoI}{\cx+0.35}
        \pgfmathsetmacro{\yoneI}{\cy-0.35}
        \pgfmathsetmacro{\ytwoI}{\cy+0.35}
        \xdef\holepaths{\holepaths
            (\xoneO, \yoneO, \goldTop) -- (\xtwoO, \yoneO, \goldTop) -- (\xtwoO, \ytwoO, \goldTop) -- (\xoneO, \ytwoO, \goldTop) -- cycle
            (\xoneI, \yoneI, \goldTop) -- (\xtwoI, \yoneI, \goldTop) -- (\xtwoI, \ytwoI, \goldTop) -- (\xoneI, \ytwoI, \goldTop) -- cycle
        }
    }
}
\fill[goldtop, even odd rule, draw=black!30, thin] (0,0,\goldTop) -- (\L,0,\goldTop) -- (\L,\L,\goldTop) -- (0,\L,\goldTop) -- cycle \holepaths;

\foreach \cx in {1.1, 3.3, 5.5} {
    \foreach \cy in {1.1, 3.3, 5.5} {
        \pgfmathsetmacro{\xoneO}{\cx-0.65}
        \pgfmathsetmacro{\xtwoO}{\cx+0.65}
        \pgfmathsetmacro{\yoneO}{\cy-0.65}
        \pgfmathsetmacro{\ytwoO}{\cy+0.65}
        \pgfmathsetmacro{\xoneI}{\cx-0.35}
        \pgfmathsetmacro{\xtwoI}{\cx+0.35}
        \pgfmathsetmacro{\yoneI}{\cy-0.35}
        \pgfmathsetmacro{\ytwoI}{\cy+0.35}

        \filldraw[fill=cyan!30, draw=cyan!70!blue, thick, opacity=0.55, even odd rule]
        (\xoneO, \yoneO, \goldTop) -- (\xtwoO, \yoneO, \goldTop) -- (\xtwoO, \ytwoO, \goldTop) -- (\xoneO, \ytwoO, \goldTop) -- cycle
        (\xoneI, \yoneI, \goldTop) -- (\xtwoI, \yoneI, \goldTop) -- (\xtwoI, \ytwoI, \goldTop) -- (\xoneI, \ytwoI, \goldTop) -- cycle;
    }
}

\draw[black!50, thick] (0, \L, \gaasBot) -- (\L, \L, \gaasBot) -- (\L, 0, \gaasBot);
\draw[black!50, thick] (\L, \L, \gaasBot) -- (\L, \L, \goldTop);
\draw[black!50, thick] (0, \L, \gaasBot) -- (0, \L, \goldTop);
\draw[black!50, thick] (\L, 0, \gaasBot) -- (\L, 0, \goldTop);

\tikzset{right face text/.style={
    font=\Large\bfseries\sffamily,
    text=black!80,
    yslant=-0.577,
    xscale=0.866,
    anchor=center
}}

\node[right face text] at (6.5, 3.3, 3.85) {Gold Metasurface};
\node[right face text] at (6.5, 3.3, 3)  {GaAs cap + AlGaAs};
\node[right face text, font=\large\bfseries\sffamily, text=red] at (6.5, 3.3, 2.25) {2DEG};
\node[right face text] at (6.5, 3.3, 1.25) {GaAs};

\end{tikzpicture}
}%
};

\node[
    anchor=north west,
    font=\bfseries\fontsize{15}{15}\selectfont
] at ([xshift=-2mm,yshift=4mm]A.north west) {(a)};
\end{tikzpicture}

\end{minipage}
\hfill
\begin{minipage}[t]{0.53\textwidth}
\centering
\vspace{0pt}

\begin{tikzpicture}
    \node[inner sep=0] (B)
    {\includegraphics[
        width=\linewidth,
        trim={0mm 0mm 0mm 0mm},
        clip
    ]{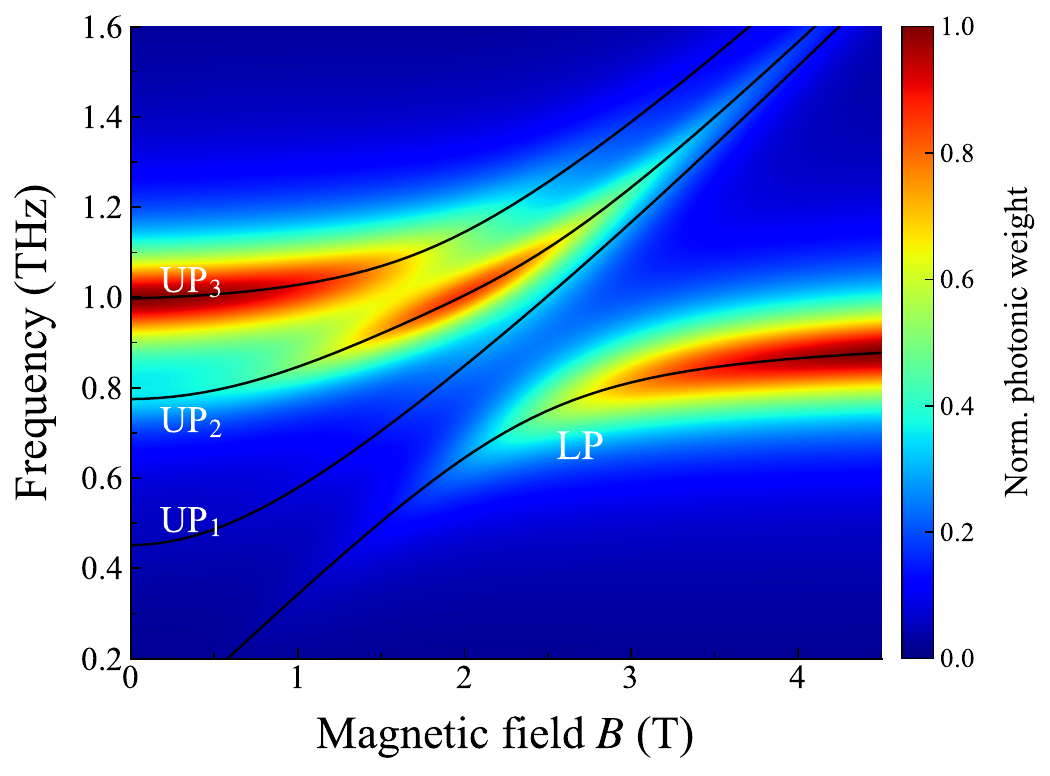}};

    \node[
        anchor=north west,
        font=\bfseries\fontsize{15}{15}\selectfont
    ] at ([xshift=-2mm,yshift=4mm]B.north west) {(b)};
\end{tikzpicture}

\end{minipage}

\caption{
\textbf{Model system used in this work, previously demonstrated experimentally.}
\textbf{(a)} Schematic of the terahertz Landau polariton device: gold nanoslot cavities patterned above a GaAs/AlGaAs heterostructure hosting a 2DEG.
\textbf{(b)} Lorentzian broadened photonic weight map for bright branches calculated from the multimode Hopfield model, reproducing the detected bright polariton branches. The map benchmarks the active spectroscopic sector.
}
\label{fig:platform_spectral_benchmark}
\end{figure*}

To test the symmetry prediction in a realistic setting, we use the multimode Landau polariton system of Ref.~\cite{EndoKimLiangLeeKimCovarrubiasMoralesSeoManfraLeeBambaKono+2025+4647+4654}, whose polarization asymmetric Hamiltonian was experimentally inferred from transmission spectroscopy. The system consists of an array of gold slot cavities fabricated on a GaAs heterostructure hosting a high-mobility two-dimensional electron gas (2DEG), as sketched in Fig.~\ref{fig:platform_spectral_benchmark}a. The subwavelength confinement of the slot mode provides two essential ingredients: strong overlap between the THz near field and the 2DEG, and finite in-plane momentum components that allow the same cavity resonance to couple both to the local \(k=0\) cyclotron resonance  (CR) and to finite-\(k\) magnetoplasmons. Such architectures build on the broader development of THz metamaterial and metasurface resonators, where strongly confined near fields enable ultrastrong light--matter coupling in deeply subwavelength mode volumes~\cite{ScalariScience2012,PhysRevB.90.205309,RajabaliNatCommun2022}.

For the device parameters considered here, the bare cavity frequency is \(\omega_0/2\pi\simeq0.925~\mathrm{THz}\). Among nonlocal modes, only modes with odd in-plane momentum components \(k=n\pi/d\) (\(n=1,3,\dots\)) are selectively excited due to the finite slot width \(d=4~\mu\mathrm{m}\). The experimentally resolved multimode structure is captured by retaining the two lowest finite-\(k\) components, MP\(_1\) and MP\(_3\), together with the CR. A perpendicular magnetic field tunes the matter spectrum through \(\omega_c=eB/m^\ast\) where \(m^\ast=0.076\,m_0\) and
\begin{equation}
\omega_{\mathrm{MP},n}^2(B)=\omega_{p,n}^2+\omega_c^2(B),
\end{equation}
producing distinct cavity detunings for CR, MP\(_1\), and MP\(_3\).  The bare mode basis, active chiral block, and numerical parameter set used in these calculations are summarized in Supplementary Note 1.
The normalized coupling strengths extracted from the measured branch positions are \(g_{\mathrm{CR}}/\omega_0\simeq0.18\) for the CR channel and \(g_1/\omega_0\simeq g_3/\omega_0\simeq0.084\) for the magnetoplasmons. Thus the CR channel lies in the ultrastrong coupling regime, while the finite-\(k\) modes form a strongly coupled multimode sector.

The corresponding polarization-resolved multimode Hopfield Hamiltonian is now expanded with the cyclotron mode \(b\) and the magnetoplasmons \(c_n\):
\begin{align}
H &= \omega_0(a_+^\dagger a_+ + a_-^\dagger a_-) + \omega_c(B)\, b^\dagger b
+ \sum_{n=1,3}\omega_{\mathrm{MP},n}(B)\, c_n^\dagger c_n
\nonumber\\
&\quad + i\Big[\Big(\bar g_{\mathrm{CR}}\, b^\dagger + \sum_{n=1,3}\bar g_n\, c_n^\dagger\Big)(a_+ + a_-^\dagger)
\nonumber\\
&\quad\ - \Big(\bar g_{\mathrm{CR}}\, b + \sum_{n=1,3}\bar g_n\, c_n\Big)(a_- + a_+^\dagger)\Big]
\nonumber\\
&\quad + D\,(a_- + a_+^\dagger)(a_+ + a_-^\dagger),
\label{eq:H_platform_results}
\end{align}
with
\begin{align}
\bar g_{\mathrm{CR}}(B)&=g_{\mathrm{CR}}\sqrt{\frac{\omega_c(B)}{\omega_0}}, \qquad
\bar g_n(B)=g_n\sqrt{\frac{\omega_{\mathrm{MP},n}(B)}{\omega_0}},
\end{align}
and
\begin{align}
 D=&\frac{\bar g_{\mathrm{CR}}^2}{\omega_c}
+\sum_{n=1,3} \frac{\bar g_n^2}{\omega_{\mathrm{MP},n}}.
\end{align}

Figure~\ref{fig:platform_spectral_benchmark}b is the Lorentzian broadened photonic weight~\cite{PhysRevB.98.205301,CiutiPRB2005} map as a spectroscopic benchmark for this model (as detailed in the Supplementary Note 3). The photonic weight map visualizing the active sector of the Hopfield Hamiltonian reproduces the experimentally observed branch pattern with the sequential avoided crossings. With this excellent agreement with the experimental spectrum established, we proceed to investigate the equilibrium vacuum correlations of the same Hamiltonian, using the chiral selection rule and quantum information diagnostics. If the symmetry prediction is correct, the polarization that organizes the observed anticrossings should remain weakly correlated with matter in the ground state, whereas the orthogonal counter-rotating polarization should carry the dominant squeezing and entanglement channels.

\paragraph*{\textbf{Vacuum entanglement routing through the inactive polarization.}}

We quantified the degree of entanglement using the logarithmic negativity measure \(E_{\mathcal N}\); the definitions and supporting photon--matter diagnostics are given in Supplementary Note 5. Figure~\ref{fig:symmetry}a confirms the central symmetry prediction. Although the active polarization \(a_+\) organizes the observed polariton anticrossings, its bipartite ground state entanglement with all matter excitations remains indistinguishable from zero within numerical precision over the entire magnetic field range. In contrast, the inactive polarization \(a_-\) exhibits finite logarithmic negativities with the CR and MP modes. That is,
\begin{equation}
E_{\mathcal N}(a_+:\mathrm{CR}\ or\ \mathrm{MP}_1\ or\ \mathrm{MP}_3)= 0,
\end{equation}
whereas
\begin{equation}
E_{\mathcal N}(a_-:\mathrm{CR}\ or\ \mathrm{MP}_1\ or\ \mathrm{MP}_3) > 0.
\end{equation}
The quantum information analysis therefore verifies the symmetry-based prediction: while the active polarization \(a_+\) remains effectively separable from the matter subsystem in the ground state, the inactive polarization \(a_-\) is precisely the channel through which entanglement and equilibrium squeezing are distributed.

\begin{figure*}[t]
\centering

\begin{minipage}[t]{0.49\textwidth}
\centering
\begin{tikzpicture}
    \node[inner sep=0] (A)
    {\includegraphics[
        width=\linewidth,
        trim={0mm 0mm 0mm 0mm},
        clip
    ]{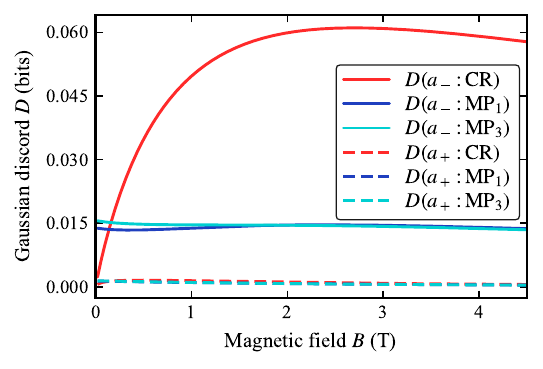}};

    \node[
        anchor=north west,
        font=\bfseries\fontsize{15}{15}\selectfont
    ] at ([xshift=0mm,yshift=3mm]A.north west) {(a)};
\end{tikzpicture}
\end{minipage}
\hfill
\begin{minipage}[t]{0.49\textwidth}
\centering
\begin{tikzpicture}
    \node[inner sep=0] (B)
    {\includegraphics[
        width=\linewidth,
        trim={0mm 0mm 0mm 0mm},
        clip
    ]{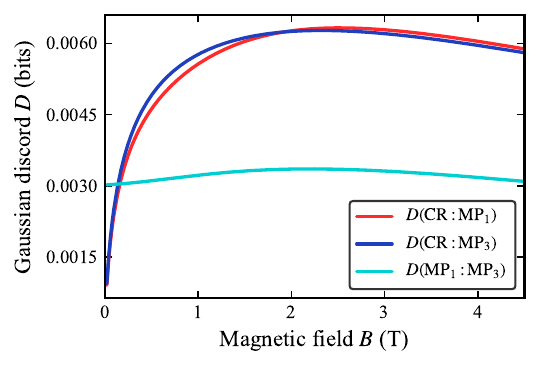}};

    \node[
        anchor=north west,
        font=\bfseries\fontsize{15}{15}\selectfont
    ] at ([xshift=0mm,yshift=3mm]B.north west) {(b)};
\end{tikzpicture}
\end{minipage}

\caption{
\textbf{Cavity-mediated nonclassical correlations in the vacuum graph state.}
\textbf{(a)} Gaussian discord between each cavity polarization and the matter modes as a function of magnetic field. The inactive polarization \(a_-\) develops substantially stronger nonclassical correlations with the cyclotron and magnetoplasmon modes than the spectroscopically active polarization \(a_+\), in line with the trend observed in the logarithmic negativity \(E_{\mathcal N}\).
\textbf{(b)} Gaussian discord, between the matter modes. Although pairwise matter--matter entanglement vanishes throughout the field range, finite discord persists between the cyclotron and magnetoplasmon excitations. This shows that the hidden \(a_-\)-centered vacuum sector leaves a nonclassical but non-entangling footprint inside the electronic subsystem. Gaussian discord is defined via Eq. 35 of Supplementary Note 5.
}
\label{fig:discord_structure}
\end{figure*}

Among the three inactive sector pairs, the strongest entanglement is found in the \(a_-\)--CR channel. It increases rapidly with magnetic field and reaches a broad maximum near the cyclotron zero detuning point. The correlations \(E_{\mathcal N}(a_-:\mathrm{MP}_1)\) and \(E_{\mathcal N}(a_-:\mathrm{MP}_3)\) are weaker and vary more gently with field, but remain clearly above zero across the investigated interval. This hierarchy reflects the coupling strengths and detunings in the system.

The low field limit further clarifies the origin of this behavior. As \(B\to 0\), the cyclotron frequency \(\omega_c=eB/m^\ast\) and the associated field-dependent coupling \(\bar g_{\mathrm{CR}}\propto \sqrt{\omega_c/\omega_0}\) both vanish. Consequently, the entanglement \(E_{\mathcal N}(a_-:\mathrm{CR})\) tends to zero in the same limit. The magnetoplasmon channels behave differently. Their frequencies satisfy
\begin{equation}
\omega_{\mathrm{MP},n}^2(B)=\omega_{p,n}^2+\omega_c^2(B),
\end{equation}
so that, as the magnetic field is reduced, they continuously approach the underlying two-dimensional plasmon modes rather than disappearing. This difference is directly visible in the bare dispersions, shown as dashed lines in the left panel of Fig. ~\ref{fig:symmetry}a. Accordingly, the inactive sector entanglement involving MP\(_1\) and MP\(_3\) survives down to low magnetic fields, while the CR-related entanglement vanishes together with the cyclotron mode.

\paragraph*{\textbf{Cavity-mediated quantum correlations across local and nonlocal matter modes.}}

The vanishing of bipartite entanglement between \(a_+\) and the matter modes does not, by itself, exclude all quantum correlations in that sector. In general, entanglement does not represent all forms of quantumness: even in the absence of entanglement, separable states can still possess quantum correlations, captured quantum discord~\cite{PhysRevLett.88.017901,LHenderson_2001}. For continuous variable Gaussian states, the corresponding measure is Gaussian discord~\cite{PhysRevLett.105.030501,GiordaParis2010}. We therefore first examine whether the polarization hierarchy found from logarithmic negativity persists at the level of Gaussian discord. Because Gaussian discord depends on which subsystem is measured, we use its symmetrized form, as defined in Eq.~35 of Supplementary Note 5.

Figure~\ref{fig:discord_structure}a shows that the Gaussian discord follows the same hierarchy already observed in the logarithmic negativity. The symmetry selected \(a_-\) sector carries the dominant Gaussian discord with the matter modes, while the corresponding \(a_+\)--matter discord remains negligible. The strongest contribution again occurs in the \(a_-\)--CR channel, while the \(a_-\)--MP correlations are weaker but finite.

Having established that the relevant cavity channel for quantum correlations is \(a_-\), we next ask what kind of correlations this hidden channel leaves inside the matter subsystem itself. There is growing interest in using cavity fields not only to hybridize individual excitations spectroscopically, but also to mediate effective interactions and correlations between distinct matter modes~\cite{GarciaVidalCiutiEbbesen2021,SchaferFlickRoncaNarangRubio2022,Kim2025Multimode,Tay2025Multimode}. Cavity-mediated interactions reshape material properties in the multimode ultrastrong coupling regime~\cite{LiuSciAdv2025}. In Landau polariton systems, this question of cavity mediation is particularly compelling because the cavity can couple both to the local \(k=0\) cyclotron resonance and to nonlocal finite-\(k\) magnetoplasmons.

\begin{figure*}[t]
\centering

\begin{minipage}[t]{0.49\textwidth}
\centering
\begin{tikzpicture}
    \node[inner sep=0] (A)
    {\includegraphics[width=\linewidth]{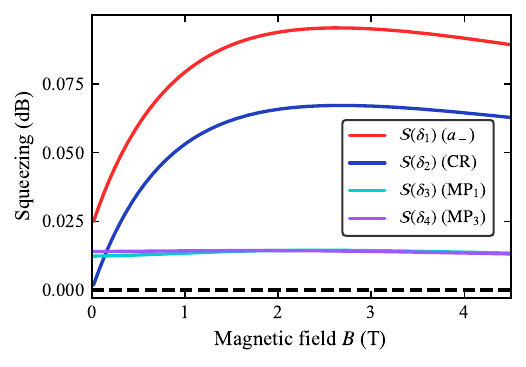}};
    \node[
        anchor=north west,
        font=\bfseries\fontsize{15}{15}\selectfont
    ] at ([xshift=0mm,yshift=2mm]A.north west) {(a)};
\end{tikzpicture}
\end{minipage}
\hfill
\begin{minipage}[t]{0.50\textwidth}
\centering
\begin{tikzpicture}
    \node[inner sep=0] (B)
    {\includegraphics[width=\linewidth]{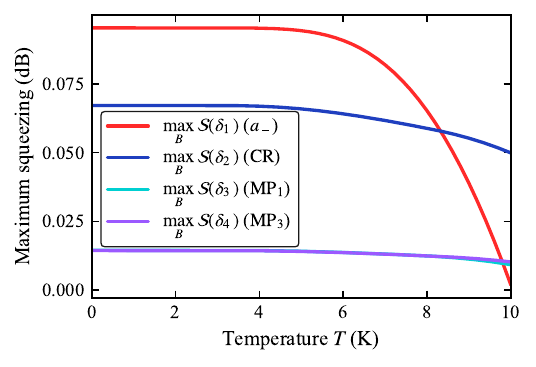}};
    \node[
        anchor=north west,
        font=\bfseries\fontsize{15}{15}\selectfont
    ] at ([xshift=0mm,yshift=3mm]B.north west) {(b)};
\end{tikzpicture}
\end{minipage}

\caption{
\textbf{Nullifier squeezing and thermal robustness of the vacuum graph state.}
\textbf{(a)} Nullifier squeezing, Eq.~39 of Supplementary Note 7, of the reduced four mode sector
\(\{a_-,\mathrm{CR},\mathrm{MP}_1,\mathrm{MP}_3\}\), showing simultaneous
subvacuum suppression of the optimized graph state nullifiers.
\textbf{(b)} Thermal robustness of the vacuum correlation structure.
The maximum nullifier squeezing remains essentially unchanged in the cryogenic regime and decreases only when thermal populations become appreciable.
}
\label{fig:graph_vacuum_sector}
\end{figure*}

No detectable pairwise entanglement emerges between any pair of matter modes across the investigated magnetic field range, even though the cavity redistributes vacuum entanglement predominantly into the \(a_-\) sector (as detailed in Supplementary Note 5). Nevertheless, Fig.~\ref{fig:discord_structure}b shows that all matter pairs exhibit finite Gaussian discord throughout the magnetic field range. Correlations involving the cyclotron resonance are the strongest, while the MP\(_1\)--MP\(_3\) pair remains weaker but nonclassically correlated. The cavity therefore imprints a quantum correlation structure inside the matter subsystem, but in a nontrivial form beyond pairwise entanglement: the matter modes remain separable, yet they are nonclassically correlated. Together with the finite \(a_-\)--matter correlations in Fig.~\ref{fig:discord_structure}a, this correlation hierarchy points to a collective \(a_-\)-centered correlated sector rather than to independent matter--matter pairs.

\paragraph*{\textbf{A nearly isolated \(a_-\)-centered sector hosting graph-like vacuum correlations.}}

The hierarchy established above suggests that the relevant correlated object is the reduced sector \(\{a_-,\mathrm{CR},\mathrm{MP}_1,\mathrm{MP}_3\}\). This sector behaves as a nearly isolated correlated subsystem, as indicated by the high purity of the reduced state \(\rho_{a_-,\mathrm{CR},\mathrm{MP}_1,\mathrm{MP}_3}\). As shown in Fig.~6c of Supplementary Note 5, the purity remains extremely close to unity throughout the full magnetic field range, with \(\mu>0.999\) for the present parameter set. The spectroscopically active polarization \(a_+\) is only weakly connected to the dominant correlated sector, so tracing it out introduces very little mixedness. This point is important conceptually as inseparability across all bipartitions certifies genuine multipartite entanglement in the pure state limit~\cite{GUHNE20091}. Here the reduced state is not strictly pure, but its near unity purity allows the multipartite structure evidenced by the bipartition analysis in Supplementary Note 7 to be effectively interpreted as a genuinely multipartite entangled sector.

Having established the near isolation and multipartite character of this sector, we next ask whether its correlations possess an organized graph-like form. We therefore examine the continuous variable nullifiers, Eq.~39 of Supplementary Note 7, constructed from the effective adjacency matrix of the reduced covariance matrix~\cite{MenicucciPRL2006,ZhangBraunstein2006,MenicucciPRA2011}. The construction of the effective adjacency matrix and the optimized nullifier variances is detailed in Supplementary Note 7. Figure~\ref{fig:graph_vacuum_sector}a shows that all optimized nullifiers exhibit positive squeezing in dB throughout the magnetic field range. Equivalently, their variances remain below the vacuum limit of \(1/2\), demonstrating simultaneous subvacuum suppression in the linear combinations that define the effective graph structure. The strongest squeezing is associated with \(a_-\) and the cyclotron mode, while the magnetoplasmon nullifiers exhibit weaker but still systematic suppression, reproducing the hierarchy already visible in the correlation measures.

These results support the graph-like interpretation anticipated schematically in Fig.~\ref{fig:symmetry}c. Because the reduced state is nearly pure and its nullifiers are simultaneously squeezed, the covariance matrix is consistent with an effective Gaussian graph state description in which \(a_-\) acts as the central hub, while the matter excitations form the surrounding nodes. The extracted adjacency matrix shown in Fig.~8c of Supplementary Note 7 reflects the same structure, with the strongest edge associated with the \(a_-\)--CR connection. In this sense, the structured vacuum correlations are neither random nor merely spectroscopic by-products of hybridization; they organize into a coherent multipartite sector selected by the inactive polarization.

Taken together, the purity, bipartition structure, and nullifier squeezing identify \(\{a_-,\mathrm{CR},\mathrm{MP}_1,\mathrm{MP}_3\}\) as a nearly isolated \(a_-\)-centered sector with effective multipartite entanglement and an internal graph-like organization.

\paragraph*{\textbf{Thermal robustness in the cryogenic regime.}}

\begin{figure*}[t]
\centering

\begin{minipage}[t]{0.48\textwidth}
\centering
\begin{tikzpicture}
    \node[inner sep=0] (A)
    {\includegraphics[
        width=\linewidth,
        trim={0mm 0mm 0mm 0mm},
        clip
    ]{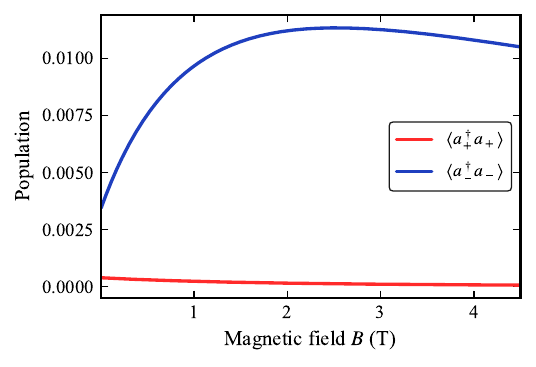}};
    \node[
        anchor=north west,
        font=\bfseries\fontsize{15}{15}\selectfont
    ] at ([xshift=0mm,yshift=5mm]A.north west) {(a)};
\end{tikzpicture}
\end{minipage}
\hfill
\begin{minipage}[t]{0.48\textwidth}
\centering
\begin{tikzpicture}
    \node[inner sep=0] (B)
    {\includegraphics[
        width=\linewidth,
        trim={0mm 0mm 0mm 0mm},
        clip
    ]{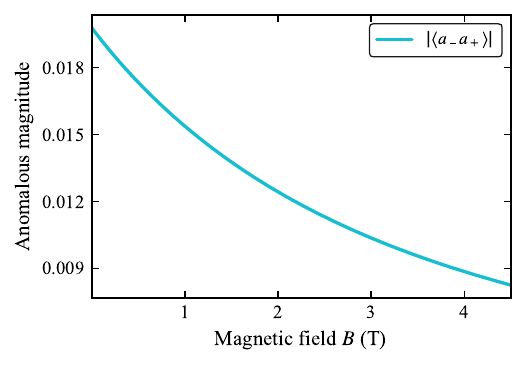}};
    \node[
        anchor=north west,
        font=\bfseries\fontsize{15}{15}\selectfont
    ] at ([xshift=-3mm,yshift=4mm]B.north west) {(b)};
\end{tikzpicture}
\end{minipage}

\vspace{2.5mm}

\begin{minipage}[t]{0.48\textwidth}
\centering
\begin{tikzpicture}
    \node[inner sep=0] (C)
    {\includegraphics[
        width=\linewidth,
        trim={0mm 0mm 0mm 0mm},
        clip
    ]{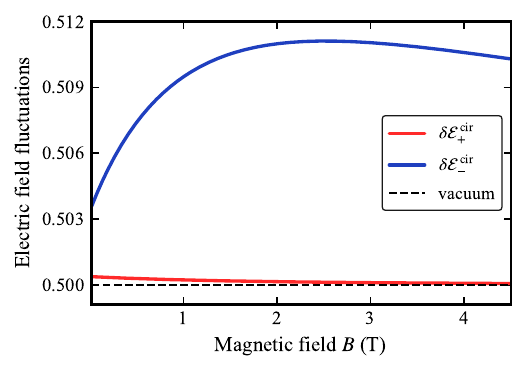}};
    \node[
        anchor=north west,
        font=\bfseries\fontsize{15}{15}\selectfont
    ] at ([xshift=-3mm,yshift=8mm]C.north west) {(c)};
\end{tikzpicture}
\end{minipage}
\hfill
\begin{minipage}[t]{0.5\textwidth}
\centering
\begin{tikzpicture}
    \node[inner sep=0] (D)
    {\includegraphics[
        width=\linewidth,
        trim={0mm 0mm 0mm 0mm},
        clip
    ]{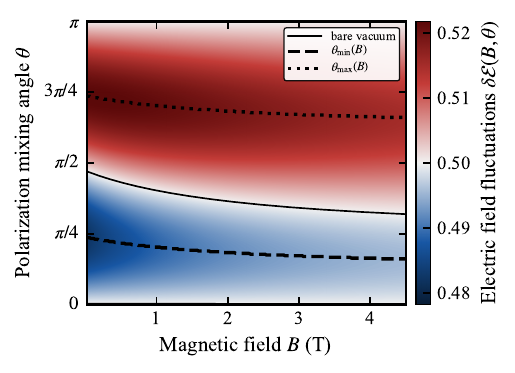}};
    \node[
        anchor=north west,
        font=\bfseries\fontsize{15}{15}\selectfont
    ] at ([xshift=-3mm,yshift=4mm]D.north west) {(d)};
\end{tikzpicture}
\end{minipage}

\caption{
\textbf{Polarization anisotropy of dressed vacuum field fluctuations.}
\textbf{(a)} Photonic populations in the dressed ground state, showing the unequal dressing of the two circular cavity modes.
\textbf{(b)} Magnitude of the anomalous correlation between the circular modes, which contributes to the mixing angle dependence of field fluctuations.
\textbf{(c)} Circular basis electric field fluctuations, Eq.~\eqref{delta_E_circ}, relative to the bare vacuum level. The active circular component remains close to the vacuum value, whereas the orthogonal component develops enhanced dressed vacuum fluctuations.
\textbf{(d)} Electric field fluctuation $\delta \mathcal{E}(B,\theta)$, Eq.~\eqref{eq:deltaVE}, map as a function of magnetic field and polarization mixing angle.
The black solid line marks the bare vacuum level, while the dashed and dotted curves indicate the polarization mixing angles of minimum and maximum fluctuation. The resulting anisotropy provides a field level signature of the chiral routing of vacuum correlations into the \(a_-\)-dominated sector.
}
\label{fig:vacuum_anisotropy}
\end{figure*}
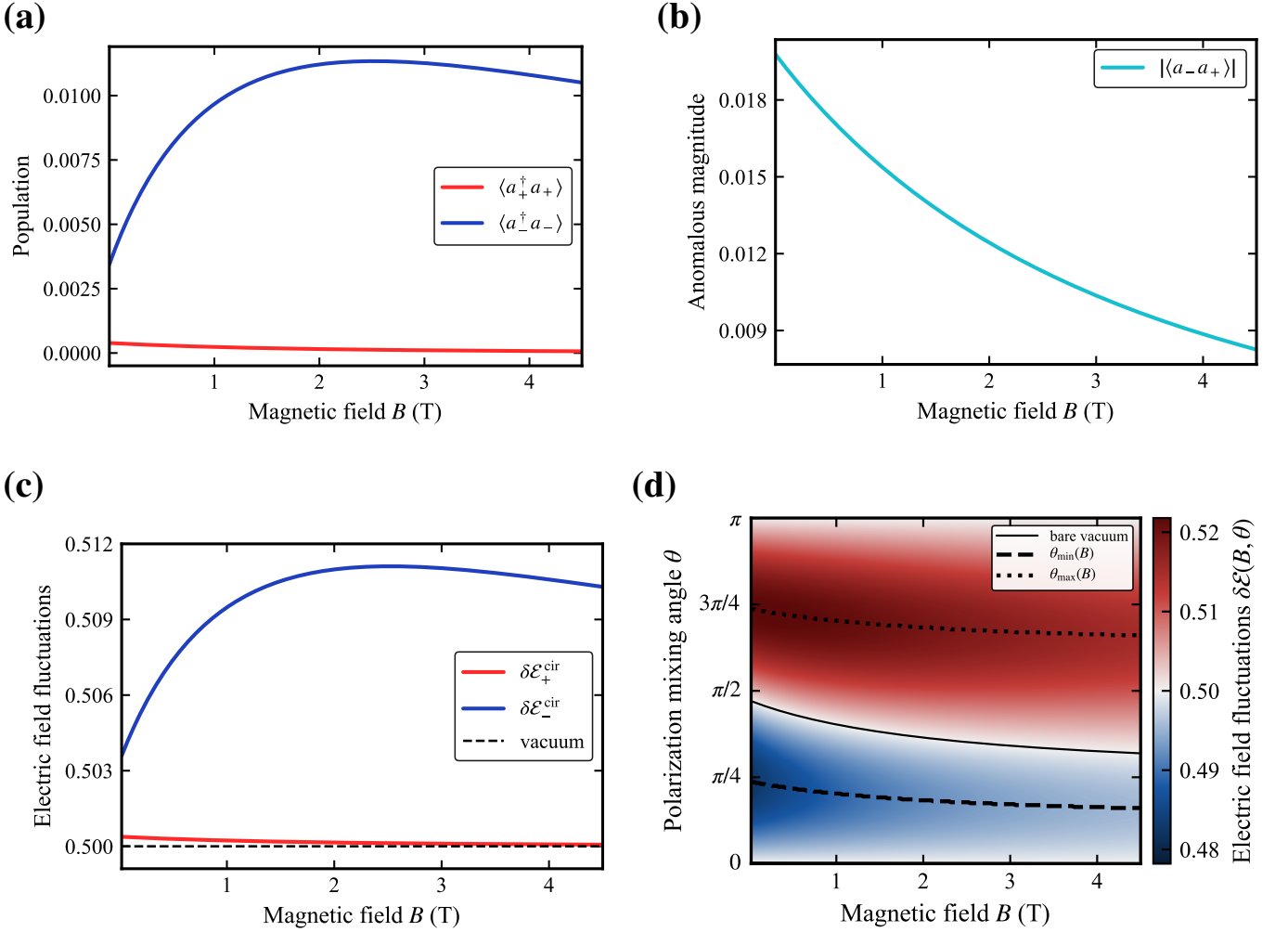

The four mode vacuum correlation structure remains robust throughout the cryogenic temperature window typically explored in Landau polariton experiments~\cite{ScalariScience2012,ZhangNatPhys2016,ParaviciniBagliani2019}. As shown in Fig.~\ref{fig:graph_vacuum_sector}b, the maximum nullifier squeezing is essentially unchanged up to about \(4\text{--}5~\mathrm{K}\), including the \(1.6~\mathrm{K}\) experimental condition of the device considered here~\cite{EndoKimLiangLeeKimCovarrubiasMoralesSeoManfraLeeBambaKono+2025+4647+4654}. At higher temperatures, thermal populations progressively wash out the anomalous correlations that underlie the graph-like vacuum structure, and the nullifier squeezing decreases accordingly. 

This reduction is not uniform across the four nodes: the squeezing associated with \(a_-\) drops faster than the others. Such behavior is consistent with the finite temperature form of the anomalous correlators, in which Bose Einstein occupations dress the pairing channels responsible for the ground state squeezing. Because the dominant anomalous structure of the correlated sector is routed through \(a_-\), its nullifier is affected by thermal admixture through the largest set of relevant channels. The centrality that makes \(a_-\) the strongest squeezed node in the ground state therefore also makes it the most thermally fragile component of the four mode structure.

\paragraph*{\textbf{Polarization resolved vacuum field fluctuations revealing the chiral vacuum structure.}}

Having established that chiral symmetry routes the dominant anomalous correlations into the inactive polarization sector, we now ask how the same symmetry shapes the cavity vacuum electric field. We define the Hermitian electric field quadratures associated with the two circular cavity modes, following~\cite{Li2018}, as
\begin{equation}
E_\pm^{\mathrm{cir}} = i E_0 \bigl(a_\pm^\dagger - a_\pm\bigr),
\end{equation}
where \(E_0\) denotes the single-mode vacuum electric-field amplitude at the evaluation point, determined by the cavity mode profile. Then, the dressed vacuum fluctuations of each circular component can be characterized by
\begin{equation}
\delta \mathcal{E}_\pm^{\mathrm{cir}}(B)
=
\frac{\langle (\Delta E_\pm^{\mathrm{cir}})^2 \rangle}{2E_0^2},
\end{equation}
which reduces to the shot-noise limit of $1/2$ in the uncoupled bare vacuum limit.

The circular basis already exposes the chiral hierarchy. Because self-anomalous correlators are forbidden by the chiral selection rule, the variance of each circular field reduces to
\begin{equation}
\delta \mathcal{E}_\pm^{\mathrm{cir}}
= \frac{1}{2} + \langle a_\pm^\dagger a_\pm \rangle,
\label{delta_E_circ}
\end{equation}
so that \(\delta \mathcal{E}_\pm^{\mathrm{cir}}\) directly tracks the corresponding virtual photon population \(n_\pm = \langle a_\pm^\dagger a_\pm \rangle\). The active circular component then remains essentially at the bare vacuum level because \(n_+\) is negligible, whereas the orthogonal polarization develops enhanced fluctuations with increasing magnetic field due to the growth of \(n_-\), as shown in panel (a) and (c) of Fig.~\ref{fig:vacuum_anisotropy}. In this way, circular electric fields reflect the asymmetric dressing of the two circular cavity components enforced by chirality.

A general polarization-resolved field measurement, however, does not need to align with either circular component. We therefore define a Hermitian quadrature parameterized by a polarization mixing angle $\theta$
\begin{equation}
E_\theta = \cos\theta\, E_+^{\mathrm{cir}} + \sin\theta\, E_-^{\mathrm{cir}},
\end{equation}
and its normalized vacuum fluctuations
\begin{equation}
\delta \mathcal{E}(B,\theta)
=
\frac{\langle (\Delta E_\theta)^2 \rangle}{2E_0^2}.
\label{eq:deltaVE}
\end{equation}
In terms of the underlying mode operators, this variance contains the two circular population contributions and an interference term,
\begin{equation}
\delta \mathcal{E}(B,\theta)
= \frac{1}{2}+
\cos^2\theta\, n_+
+
\sin^2\theta\, n_-
-
\sin(2\theta)\,\mathrm{Re}\,\langle a_+ a_- \rangle.
\end{equation}
The first three terms describe the enhancement of field fluctuations above the bare vacuum level of $1/2$, arising from the presence of virtual populations $n_\pm$. In this way, vacuum fluctuations provide an indirect probe of virtual particles through measurable noisy field signatures. The last term, proportional to the anomalous correlator $\langle a_+ a_- \rangle$, represents an interference contribution that can suppress the fluctuations and even squeeze them below the shot-noise limit. This interference between the two circular components fades with increasing magnetic field (panel (b) of Fig.~\ref{fig:vacuum_anisotropy}). Both effects depend strongly on the magnetic field B, which controls the virtual populations and the strength of the anomalous correlations, and on the polarization mixing angle $\theta$, which reflects the anisotropic nature of the dressed vacuum. If the cavity vacuum were trivial, \textit{i.e.}, in the absence of virtual populations and anomalous correlations, the field fluctuations would remain at the bare vacuum level, yielding $\delta \mathcal{E}(B,\theta) = 1/2$ for all polarization mixing. Therefore, the robust angular dependence of Eq.~\eqref{eq:deltaVE} shown in Figure~\ref{fig:vacuum_anisotropy}d, is not an independent effect on top of the correlation analysis, but the field level manifestation of the same chiral routing mechanism.

\section{Discussion}

We showed that chiral ultrastrong coupling separates two roles of the cavity that are often implicitly associated with one another. In the Landau polariton system studied here, the polarization \(a_+\) remains the spectroscopically bright channel that organizes the avoided crossings, while the dominant anomalous correlators, vacuum squeezing, and cavity--matter entanglement are routed into the orthogonal polarization \(a_-\). The polarization that forms the observed polariton branches is not necessarily the one that carries the nonclassical structure of the ground state~\cite{Li2018,EndoKimLiangLeeKimCovarrubiasMoralesSeoManfraLeeBambaKono+2025+4647+4654,PhysRevLett.123.133603}. This distinction is essential for controlling cavity of matter as recent theory has emphasized that light--matter entanglement can be an essential ingredient through which cavity quantum materials can modify some properties via vacuum field engineering~\cite{PhysRevLett.131.023601,GarciaVidalCiutiEbbesen2021,10.1063/5.0083825,Lu:25}. In Landau polariton systems, cavity vacuum fields are already known to influence equilibrium transport and collective properties of a two-dimensional electron gas~\cite{PhysRevB.98.205301,ParaviciniBagliani2019,Appugliese2022,Enkner2025}. The implication of our results is that spectroscopy and quantum information diagnostics answer complementary questions: the former determines mode hybridization, while the latter reveals which light and matter degrees of freedom share nonclassical correlations in the dressed vacuum.

The polarization-resolved field fluctuation anisotropy gives this hidden structure a signature at the field level. In the circular basis, the field fluctuation of each component directly tracks the corresponding virtual population because same-polarization anomalous moments are forbidden by the chiral symmetry. In a general mixed polarization, the anomalous correlation between the two circular components enters as an interference term, enhancing or suppressing the dressed vacuum noise depending on the mixing angle. The vacuum noise therefore acquires an angular texture set by the chiral routing of correlations. This connects naturally to electro-optic measurements of electromagnetic vacuum field correlations in the terahertz range, which have shown that vacuum field correlations can be accessed without relying on photon emission or intensity detection~\cite{BeneaChelmusNature2019}. In this sense, the predicted anisotropy serves as an experimentally accessible fingerprint of the hidden non-polaritonic vacuum sector.

The hidden correlated sector has a well defined internal organization. The mode \(a_-\) acts as the hub of the vacuum structure: it carries the strongest cavity--matter entanglement, mediates finite Gaussian discord between cyclotron and magnetoplasmon excitations, and forms, together with \(\{\mathrm{CR},\mathrm{MP}_1,\mathrm{MP}_3\}\), a nearly pure state consistent with an effective Gaussian graph state description~\cite{MenicucciPRA2011,MenicucciPRL2006,PhysRevResearch.5.033136,PhysRevA.111.052437}. By contrast, \(a_+\) is only weakly and indirectly connected to this structure. The cavity therefore does more than produce avoided crossings: it redistributes nonclassical correlations across distinct collective channels of the 2DEG. In particular, the finite matter--matter quantum discord shows that local cyclotron and nonlocal magnetoplasmon modes can remain quantum correlated even when they are not pairwise entangled. This identifies Gaussian discord as an indicator of the hidden vacuum structure of the cavity.

More broadly, our findings point to a passive route toward structured correlated vacua in solid state cavity QED. In contrast to optical continuous variable graph states generated by external driving or nonlinear state preparation~\cite{PhysRevLett.101.130501,PhysRevA.76.010302}, the graph-like multimode Gaussian structure identified here emerges directly from symmetry and counter-rotating interactions in equilibrium. It complements recent studies of ground state entanglement and nonclassical ultrastrong-coupling vacua in bosonic and polaritonic settings~\cite{PhysRevResearch.5.033136,PhysRevA.111.052437}, but adds an organizing principle based on symmetry: chirality determines not only which excitations are bright, but also how the vacuum correlations are routed.

This perspective suggests several directions. First, polarization-resolved field correlation or response measurements could provide direct tests of the predicted asymmetry between the bright spectroscopic channel and the hidden vacuum correlation channel. Second, the matter sector discord found here motivates further study of whether non-entangling quantum correlations can influence equilibrium electronic response, especially in settings where cavity vacuum fields already modify transport. Third, the same routing principle may be engineered beyond Landau polaritons. In superconducting circuits, ingredients related to chirality and synthetic angular-momentum structure have already been demonstrated in few qubit architectures~\cite{Roushan2017}. Atomic platforms likewise offer a high degree of control over motional and spin--motion degrees of freedom~\cite{PhysRevLett.121.253603,PhysRevLett.76.1796}, and chiral quantum-optical couplings have been proposed more explicitly in Rydberg and trapped-ion settings~\cite{PhysRevA.93.063830}.

In conlusion, our results recast chirality as more than a selection rule for polariton spectroscopy. It is a symmetry principle for organizing the ultrastrong coupling vacuum itself. The chiral structure of Landau polaritons separates the bright polariton spectrum from the hidden nonclassical vacuum sector while imprinting a polarization anisotropy in the field fluctuations, thereby opening a route to explore how cavity-mediated quantum correlations, including correlations beyond entanglement, may enter the equilibrium physics of quantum materials.

\section{Methods \label{methods}}

\paragraph*{\textbf{Chiral charge and closed operator sectors.}}

The schematic Hamiltonian in Eq.~\eqref{eq:Hschematic_results} has a conserved chiral charge
\begin{equation}
Q=a_+^\dagger a_+ + \sum_j m_j^\dagger m_j - a_-^\dagger a_- ,
\label{eq:Q_methods}
\end{equation}
which assigns charge \(-1\) to \(a_+\) and the matter annihilation operators \(m_j\), and charge \(+1\) to \(a_-\). The corresponding $U(1)$ unitary transformation \(U(\theta)=e^{-i\theta Q}\) gives
\begin{align}
U^\dagger(\theta)a_+U(\theta)&=e^{-i\theta}a_+, \\
\quad
U^\dagger(\theta)m_jU(\theta)&=e^{-i\theta}m_j,\\
\quad
U^\dagger(\theta)a_-U(\theta)&=e^{+i\theta}a_- .
\end{align}
Every term in Eq.~\eqref{eq:Hschematic_results} is neutral under this transformation, and therefore
\begin{equation}
[H,Q]=0 .
\end{equation}

The same symmetry appears dynamically as a block decomposition of the Heisenberg equations. Direct evaluation gives
\begin{align}
i \dot a_{+} &= (\omega_0 + D)\,a_{+} + D\,a_{-}^\dagger - i \sum_j g_j m_j ,
\label{eq:eom_aplus}\\
i \dot m_j &= \omega_j m_j + i g_j(a_{+}+a_{-}^\dagger),
\label{eq:eom_mj}\\
i \dot a_{-}^\dagger &= -(\omega_0 + D)a_{-}^\dagger - D a_{+} + i \sum_j g_j m_j .
\label{eq:eom_aminusdag}
\end{align}
Thus the negative-charge operator vector
\begin{equation}
\bm{\eta}_-=(a_+,m_1,m_2,\ldots,a_-^\dagger)^T
\end{equation}
is closed under time evolution, \(i\dot{\eta}_-=K_-\bm{\eta}_-\), with
\begin{equation}
K_{-}=
\begin{pmatrix}
\omega_0+D & -i g_1 & -i g_2 & \cdots & -i g_{N_m} & D \\
i g_1 & \omega_1 & 0 & \cdots & 0 & i g_1 \\
i g_2 & 0 & \omega_2 & \cdots & 0 & i g_2 \\
\vdots & \vdots & \vdots & \ddots & \vdots & \vdots \\
i g_{N_m} & 0 & 0 & \cdots & \omega_{N_m} & i g_{N_m} \\
-D & i g_1 & i g_2 & \cdots & i g_{N_m} & -(\omega_0+D)
\end{pmatrix}.
\label{eq:Kminus}
\end{equation}
The Hermitian conjugate positive charge sector
\begin{equation}
\bm{\eta}_+=(a_+^\dagger,m_1^\dagger,m_2^\dagger,\ldots,a_-)^T
\end{equation}
is also closed, so that
\begin{equation}
i\frac{d}{dt}
\begin{pmatrix}
\bm{\eta}_-\\
\bm{\eta}_+
\end{pmatrix}
=
\begin{pmatrix}
K_- & 0\\
0 & K_+
\end{pmatrix}
\begin{pmatrix}
\bm{\eta}_-\\
\bm{\eta}_+
\end{pmatrix},
\qquad
K_+=-K_-^* .
\label{eq:blockdyn_methods}
\end{equation}

\paragraph*{\textbf{Selection rule for anomalous moments.}}

The block structure above is equivalent to a charge selection rule. Let \(O_q\) be an operator with definite chiral charge \(q\), such that
\begin{equation}
U^\dagger(\theta) O_q U(\theta)=e^{-iq\theta}O_q .
\end{equation}
Since \([H,Q]=0\), a nondegenerate ground state can be chosen as an eigenstate of \(Q\). Therefore,
\begin{align}
\langle O_q\rangle
&=\bra{\mathrm{GS}}O_q\ket{\mathrm{GS}} \nonumber\\
&=\bra{\mathrm{GS}}U^\dagger(\theta)O_qU(\theta)\ket{\mathrm{GS}} \nonumber\\
&=e^{-iq\theta}\langle O_q\rangle ,
\end{align}
for arbitrary \(\theta\). Hence \(\langle O_q\rangle=0\) unless \(q=0\).

Applying this rule to the anomalous correlators involving the spectroscopically active polarization gives
\begin{align}
q(a_+a_+) &= -2,
&
q(a_+m_j) &= -2,
&
q(a_+^\dagger m_j^\dagger) &= +2 ,
\end{align}
and therefore
\begin{equation}
\langle a_+a_+\rangle=0,
\qquad
\langle a_+m_j\rangle=0,
\qquad
\langle a_+^\dagger m_j^\dagger\rangle=0 .
\label{eq:selection_rule_methods}
\end{equation}
Thus all anomalous correlators involving \(a_+\) and matter annihilation operators vanish identically in any \(Q\)-eigenstate, including the ground state. By contrast, anomalous correlators involving the inactive polarization \(a_-\), such as \(\langle a_- m_j\rangle\) in the charge balanced sector or equivalently the corresponding quadrature correlations generated by counter rotating terms, are not forbidden by the \(U(1)\) symmetry. This is the origin of the selection rule of the vacuum correlation routing discussed in the main text.

Perturbations that mix circular polarizations, such as cavity birefringence, ellipticity, or breaking of the ideal selection rules by disorder, do not commute with \(Q\). Such terms can activate otherwise forbidden anomalous correlators involving \(a_+\) and provide a direct way to test the robustness of the chiral hierarchy.

\paragraph*{\textbf{Bogoliubov diagonalization and covariance matrix construction.}}

The chiral block decomposition allows the Bogoliubov problem to be formulated separately in the two charge sectors. In the negative charge sector,
\begin{equation}
\bm{\eta}_-=(a_+,m_1,m_2,\ldots,a_-^\dagger)^T ,
\end{equation}
we define normal mode operators
\begin{equation}
P_{\nu,-}
=
w_{\nu,+}\,a_+
+\sum_j w_{\nu,j}\,m_j
+y_{\nu,-}\,a_-^\dagger ,
\label{eq:Pminus_sector}
\end{equation}
which satisfy
\begin{equation}
[P_{\nu,-},H]=\Omega_{\nu,-}P_{\nu,-}.
\end{equation}
This sector contains the active cavity annihilation operator \(a_+\) and the matter annihilation operators \(m_j\), and therefore hosts the polariton branches observed in the active spectroscopic response. The inactive polarization enters this sector only through \(a_-^\dagger\), i.e. through the counter-rotating component.

The positive charge sector,
\begin{equation}
\bm{\eta}_+=(a_+^\dagger,m_1^\dagger,m_2^\dagger,\ldots,a_-)^T ,
\end{equation}
is diagonalized analogously by
\begin{equation}
P_{\nu,+}
=
\tilde y_{\nu,+}\,a_+^\dagger
+\sum_j \tilde y_{\nu,j}\,m_j^\dagger
+\tilde w_{\nu,-}\,a_- ,
\label{eq:Pplus_sector}
\end{equation}
with
\begin{equation}
[P_{\nu,+},H]=\Omega_{\nu,+}P_{\nu,+}.
\end{equation}
For the ideal chiral Hamiltonian considered here, the two blocks are related by Hermitian conjugation, \(K_+=-K_-^*\). Consequently, the positive frequency spectrum of the full Bogoliubov problem can be obtained from either block together with its conjugate partner. In practice, we retain all positive norm, positive frequency solutions and normalize them with the bosonic metric,
\begin{equation}
|w_{\nu,+}|^2+\sum_j |w_{\nu,j}|^2-|y_{\nu,-}|^2=1
\end{equation}
for the \(\bm{\eta}_-\) sector, and similarly for \(\bm{\eta}_+\). The quadratic Hamiltonian then takes the diagonal form
\begin{equation}
H=\sum_{\nu,s=\pm}\Omega_{\nu,s}\,
P_{\nu,s}^\dagger P_{\nu,s},
\label{eq:diag_chiral_blocks}
\end{equation}
up to the ground state energy, where only positive frequency normal modes are included.

The covariance matrix is obtained from Gaussian contractions in the normal mode basis. Since the Hamiltonian is quadratic, the ground state and the finite temperature Gibbs state are Gaussian. Therefore, all second moments of the bare operators follow from Wick contractions once the normal mode occupations are specified. At zero temperature,
\begin{equation}
\langle P_{\nu,s}^\dagger P_{\mu,s'}\rangle=0,
\qquad
\langle P_{\nu,s} P_{\mu,s'}^\dagger\rangle=\delta_{\nu\mu}\delta_{ss'} .
\end{equation}
At finite temperature,
\begin{equation}
\begin{aligned}
\langle P_{\nu,s}^\dagger P_{\mu,s'}\rangle
&= n_{\nu,s}(T)\,\delta_{\nu\mu}\delta_{ss'}, \\
\langle P_{\nu,s} P_{\mu,s'}^\dagger\rangle
&= \big[n_{\nu,s}(T)+1\big]\delta_{\nu\mu}\delta_{ss'} .
\end{aligned}
\end{equation}
where
\begin{equation}
n_{\nu,s}(T)=\frac{1}{\exp(\beta\Omega_{\nu,s})-1}.
\end{equation}
Substituting the inverse Bogoliubov transformation then gives the normal and anomalous correlators among the physical cavity and matter modes. These include, for example, \(\langle a_\xi^\dagger a_{\xi'}\rangle\), \(\langle m_i^\dagger m_j\rangle\), \(\langle a_\xi m_j\rangle\), and \(\langle a_\xi a_{\xi'}\rangle\), which are the ingredients entering the quadrature covariance matrix.

We introduce dimensionless quadratures for each physical mode,
\begin{equation}
x_\ell=\frac{o_\ell+o_\ell^\dagger}{\sqrt{2}},
\qquad
p_\ell=\frac{o_\ell-o_\ell^\dagger}{i\sqrt{2}},
\end{equation}
where
\begin{equation}
\{o_\ell\}=\{a_+,a_-,m_1,m_2,\ldots\}.
\end{equation}
Collecting them into
\begin{equation}
\bm R=(x_1,p_1,x_2,p_2,\ldots,x_N,p_N)^T ,
\end{equation}
the covariance matrix is
\begin{equation}
V_{jk}
=
\frac{1}{2}
\langle R_j R_k+R_k R_j\rangle .
\label{eq:cov_methods}
\end{equation}
The symplectic form is
\begin{equation}
\Omega=\bigoplus_{\ell=1}^{N}
\begin{pmatrix}
0&1\\
-1&0
\end{pmatrix},
\end{equation}
so that \([R_j,R_k]=i\Omega_{jk}\). In this convention the bare vacuum satisfies \(V_{\mathrm{vac}}=I/2\), and any physical covariance matrix obeys
\begin{equation}
V+\frac{i}{2}\Omega\ge 0 .
\label{V}
\end{equation}
Numerical checks of the Bogoliubov transformation and covariance matrix physicality are reported in Supplementary Note 2.

All purity, entanglement, discord, and graph state diagnostics used in the main text are obtained directly from the covariance matrix \(V\), Eq.~\eqref{V}. In practice, this reduces the analysis to standard linear algebra operations. Selecting subsets of quadratures to construct reduced covariance matrices, computing symplectic eigenvalues for entropies and mutual information, applying partial transposition to evaluate logarithmic negativities, optimizing the Gaussian conditional entropy for the Gaussian discord, and extracting the effective adjacency matrix that minimizes the nullifier variances.

\section*{Data availability}
The datasets used and/or analyzed during the current study
are available from the following publicly accessible repository  \href{https://github.com/MoorishQubit/Hopfield}{https://github.com/MoorishQubit/Hopfield}.

\section*{Code availability}
The source code to reproduce the results and all the data used to generate the plots is available in the accompanying GitHub repository \href{https://github.com/MoorishQubit/Hopfield}{https://github.com/MoorishQubit/Hopfield}. 

\section*{Acknowledgements}
Z.M. acknowledges funding from the Ministry of Economic Affairs, Labour and Tourism Baden-Württemberg in the frame of the Competence Center Quantum Computing Baden-Württemberg (project ``KQCBW25'').
H.R.’s material is based upon work supported by the Department of Energy [Basic Energy Sciences] under Award Number(s) [DE-FG02-02ER15344].

\section*{Author contributions}

A.E.-A. and Z.M. conceived the research project, developed the theoretical framework, derived the analytical results, and performed the numerical simulations. H.S. contributed to the theoretical analysis, interpretation of the results, and scientific discussions. H.R., A.E.F. and D.L. supervised the research and provided critical feedback throughout the project. All authors contributed to the writing and revision of the manuscript and approved the final version.

\section*{Competing Interests}
The authors declare no competing interests.

\bibliography{bibliography}

\clearpage
\onecolumngrid

\setcounter{section}{0}
\renewcommand{\thesection}{\arabic{section}}

\setcounter{equation}{0}
\setcounter{figure}{0}
\setcounter{table}{0}
\renewcommand{\theequation}{S\arabic{equation}}
\renewcommand{\thefigure}{S\arabic{figure}}
\renewcommand{\thetable}{S\arabic{table}}

\renewcommand{\theHequation}{supp.\arabic{equation}}
\renewcommand{\theHfigure}{supp.\arabic{figure}}
\renewcommand{\theHtable}{supp.\arabic{table}}
\renewcommand{\theHsection}{supp.\arabic{section}}

\onecolumngrid
\begin{center}
{\large \bf Supplemental Material:\\ Chirality routing non-polaritonic vacuum correlations in Landau polaritons}\\\vspace{0.3cm}
{Ayoub El-Amrani$^{1,*}$, Zakaria Mzaouali$^{2,3}$, Houssam Sabri$^{1}$, Herschel Rabitz$^{4}$, Abdelouahed El Fatimy$^{1}$, and Dukhyung Lee$^{1,\dagger}$}\\ \vspace{0.5em} { $^{1}$College of Physical Sciences and Engineering, University Mohammed VI Polytechnic, Ben Guerir, 43150, Morocco.\\ $^{2}$J\"ulich Supercomputing Centre, Institute for Advanced Simulation, Forschungszentrum J\"ulich, Wilhelm-Johnen-Stra{\ss}e, J\"ulich, 52428, Germany.\\ $^{3}$Institut f\"ur Theoretische Physik, Universit\"at T\"ubingen, Auf der Morgenstelle 14, 72076 T\"ubingen, Germany.\\
$^{4}$Department of Chemistry, Princeton University, Princeton, New Jersey 08544, USA.\\ $^{*,\dagger}$Corresponding authors: \texttt{ayoub.el-amrani@um6p.ma}, \texttt{dukhyung.lee@um6p.ma}}
\end{center}
\twocolumngrid

\newcommand{\suppnote}[2]{%
  \refstepcounter{section}%
  \section*{Supplementary Note~\thesection: #1}%
  \label{#2}%
}

\newcommand{\subsuppnote}[1]{%
\vspace{0.5em}
\subsection*{#1}
}

\suppnote{Model, basis, and numerical parameters}{supp1}

We summarizes the model conventions and numerical parameters used throughout the calculations. The goal is to make explicit the bare mode basis, the chiral active block used for the Bogoliubov diagonalization, and the experimentally extracted parameter set used to generate the figures in the main text. The calculations are based on the multimode Hopfield Hamiltonian, Eq.~\eqref{eq:H_platform_results}. The bare mode basis is
\begin{equation}
    \bm{a}
    =
    \left(
    a_{+},
    a_{-},
    b,
    c_{1},
    c_{3}
    \right)^{T},
\end{equation}
where \(a_{\pm}\) are the two circular cavity polarizations, \(b\) is the cyclotron resonance mode, and \(c_{1},c_{3}\) are the finite momentum magnetoplasmon modes selected by the cavity geometry.

The active chiral block is written in the basis
\begin{equation}
    \bm{\eta}_{-}
    =
    \left(
    a_{+},
    a_{-}^{\dagger},
    b,
    c_{1},
    c_{3}
    \right)^{T}.
\end{equation}
In this basis, the dynamical matrix is
\begin{equation}
M_{-}(B)=
\begin{pmatrix}
\omega_0+D & D & -i\bar g_{\mathrm{CR}} & -i\bar g_{1} & -i\bar g_{3}\\
-D & -\omega_0-D & i\bar g_{\mathrm{CR}} & i\bar g_{1} & i\bar g_{3}\\
i\bar g_{\mathrm{CR}} & i\bar g_{\mathrm{CR}} & \omega_c & 0 & 0\\
i\bar g_{1} & i\bar g_{1} & 0 & \omega_{\mathrm{MP},1} & 0\\
i\bar g_{3} & i\bar g_{3} & 0 & 0 & \omega_{\mathrm{MP},3}
\end{pmatrix}.
\end{equation}

The bare dispersions are
\begin{equation}
    \omega_c(B)=\frac{eB}{m^{*}},
\end{equation}
and
\begin{equation}
    \omega_{\mathrm{MP},n}(B)
    =
    \sqrt{\omega_{p,n}^{2}+\omega_c^2(B)}.
\end{equation}
The couplings are
\begin{equation}
    \bar{g}_{\mathrm{CR}}(B)
    =
    g_{\mathrm{CR}}\sqrt{\frac{\omega_c(B)}{\omega_0}},
    \qquad
    \bar{g}_{n}(B)
    =
    g_n\sqrt{\frac{\omega_{\mathrm{MP},n}(B)}{\omega_0}}.
\end{equation}
The diamagnetic coefficient is
\begin{equation}
    D(B)=
    \frac{\bar g_{\mathrm{CR}}^{\,2}}{\omega_c}
    +
    \sum_{n=1,3}
    \frac{\bar g_n^{\,2}}{\omega_{\mathrm{MP},n}}.
\end{equation}

The numerical parameters used in the calculations are summarized in Table~\ref{tab:parameters}. They are taken from the experimentally fitted multimode Landau polariton model of Ref.~\cite{EndoKimLiangLeeKimCovarrubiasMoralesSeoManfraLeeBambaKono+2025+4647+4654}.

\begin{table}[t]
\centering
\caption{
Numerical parameters used throughout the calculations.
}
\label{tab:parameters}
\begin{tabular}{lll}
\toprule
Quantity & Value & Meaning \\
\midrule
\(\omega_0/2\pi\) & \SI{0.925}{THz} & Bare cavity frequency \\
\(m^{*}\) & \(0.076m_0\) & Effective mass \\
\(B_{\mathrm{CR}}\) & \SI{2.51}{T} & CR--cavity zero detuning \\
\(B_{\mathrm{MP1}}\) & \SI{2.18}{T} & MP\(_1\)--cavity zero detuning \\
\(B_{\mathrm{MP3}}\) & \SI{1.25}{T} & MP\(_3\)--cavity zero detuning \\
\(g_{\mathrm{CR}}/\omega_0\) & \(0.18\) & CR coupling strength \\
\(g_{\mathrm{MP1}}/\omega_0\) & \(0.084\) & MP\(_1\) coupling strength \\
\(g_{\mathrm{MP3}}/\omega_0\) & \(0.084\) & MP\(_3\) coupling strength \\
\bottomrule
\end{tabular}
\end{table}

\suppnote{Bogoliubov consistency and covariance matrix physicality}{supp2}

Here we document the internal consistency checks of the bosonic Bogoliubov transformation and the physicality checks of the Gaussian covariance matrix. These tests ensure that the numerical diagonalization preserves bosonic commutation relations and produces physical covariance matrices throughout the magnetic field scan.

\paragraph*{Bogoliubov transformation.}

The Hamiltonian, Eq.~\eqref{eq:H_platform_results}, is diagonalized by a bosonic Bogoliubov transformation,
\begin{equation}
    P_{\nu}
    =
    \sum_{i}
    \left(
    W_{\nu i} a_i
    +
    Y_{\nu i} a_i^{\dagger}
    \right),
\end{equation}
where \(\nu\) labels the positive frequency normal modes and \(i\) labels the bare modes. With the row-wise numerical convention used here, preservation of bosonic commutation relations requires~\cite{COLPA1978327}
\begin{equation}
    W W^{\dagger}-Y Y^{\dagger}=I,
\end{equation}
and
\begin{equation}
    W Y^{T}-Y W^{T}=0.
\end{equation}
The corresponding inverse completeness checks are
\begin{equation}
    W^{\dagger}W-Y^{T}Y^{*}=I,
\end{equation}
and
\begin{equation}
    W^{\dagger}Y-Y^{T}W^{*}=0.
\end{equation}
The first pair of equations verifies that the normal mode operators are canonical bosons, while the second pair verifies completeness of the inverse transformation back to the bare cavity-matter operators.

The matrix residuals are quantified using both the Frobenius norm,
\begin{equation}
    \|R\|_{\mathrm F}
    =
    \sqrt{\mathrm{Tr}(R^\dagger R)},
\end{equation}
and the maximum absolute matrix element \(\max_{ij}|R_{ij}|\). The maximum deviations over the full magnetic field scan are given in Table~\ref{tab:bogoliubov}. All errors remain at the level of double precision numerical roundoff.

\begin{table}[t]
\centering
\caption{
Bogoliubov consistency checks. The table reports the maximum deviation over the full magnetic field range.
}
\label{tab:bogoliubov}
\begin{tabular}{lll}
\toprule
Quantity & Maximum deviation & Field position \\
\midrule
\(\|W W^{\dagger}-Y Y^{\dagger}-I\|_{\mathrm{F}}\) & \(2.352\times 10^{-14}\) & \SI{3.6190}{T} \\
\(\max |W W^{\dagger}-Y Y^{\dagger}-I|\) & \(1.658\times 10^{-14}\) & \SI{3.6190}{T} \\
\(\|W Y^{T}-Y W^{T}\|_{\mathrm{F}}\) & \(1.273\times 10^{-15}\) & \SI{0.2262}{T} \\
\(\max |W Y^{T}-Y W^{T}|\) & \(7.425\times 10^{-16}\) & \SI{0.2074}{T} \\
\(\|W^{\dagger}W-Y^{T}Y^{*}-I\|_{\mathrm{F}}\) & \(2.348\times 10^{-14}\) & \SI{3.6190}{T} \\
\(\max |W^{\dagger}W-Y^{T}Y^{*}-I|\) & \(1.488\times 10^{-14}\) & \SI{3.6190}{T} \\
\(\|W^{\dagger}Y-Y^{T}W^{*}\|_{\mathrm{F}}\) & \(2.200\times 10^{-15}\) & \SI{3.9939}{T} \\
\(\max |W^{\dagger}Y-Y^{T}W^{*}|\) & \(1.478\times 10^{-15}\) & \SI{3.9939}{T} \\
\bottomrule
\end{tabular}
\end{table}

\paragraph*{Covariance matrix.}

The covariance matrix is constructed from the quadrature vector
\begin{equation}
    \bm{R}
    =
    \left(
    q_1,\ldots,q_N,p_1,\ldots,p_N
    \right)^T,
\end{equation}
with
\begin{equation}
    q_i=\frac{a_i+a_i^{\dagger}}{\sqrt{2}},
    \qquad
    p_i=\frac{a_i-a_i^{\dagger}}{i\sqrt{2}}.
\end{equation}
The covariance matrix is
\begin{equation}
    V_{ij}
    =
    \frac{1}{2}
    \langle R_iR_j+R_jR_i\rangle.
\end{equation}
In this convention the vacuum covariance is \(V_{\mathrm{vac}}=I/2\).

A physical Gaussian covariance matrix must satisfy~\cite{WeedbrookRMP2012}
\begin{equation}
    V+\frac{i}{2}\Omega \geq 0,
\end{equation}
where
\begin{equation}
    \Omega=
    \begin{pmatrix}
    0 & I\\
    -I & 0
    \end{pmatrix}.
\end{equation}
Equivalently, all symplectic eigenvalues must satisfy
\begin{equation}
    \nu_k \geq \frac{1}{2}.
\end{equation}


\suppnote{Spectroscopic sector, photonic weights, and broadened visibility map}{supp3}

We detail here how the active sector polariton branches are obtained and how the transmission-like map used in the main text is constructed. The key point is that the map is not a full transfer matrix or input--output transmission calculation. It is a Lorentzian broadened photonic weight visibility proxy that visualizes the bright $a_+$-sector of the Hopfield model.

\begin{figure}[t]
\centering
\includegraphics[width=\linewidth]{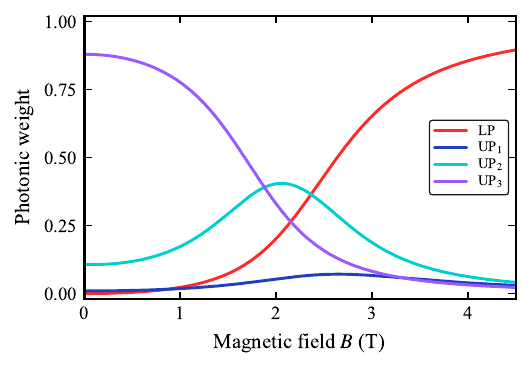}
\caption{
Net photonic weights \(V_\nu\), Eq.~\eqref{eq:photonic_weight}, by branch. These weights quantify the spectroscopic visibility of the active polariton branches and provide the input weights for the broadened map.
}
\label{fig:supp_photonic_weights}
\end{figure}

\begin{figure}[t]
\centering
\includegraphics[width=\linewidth]{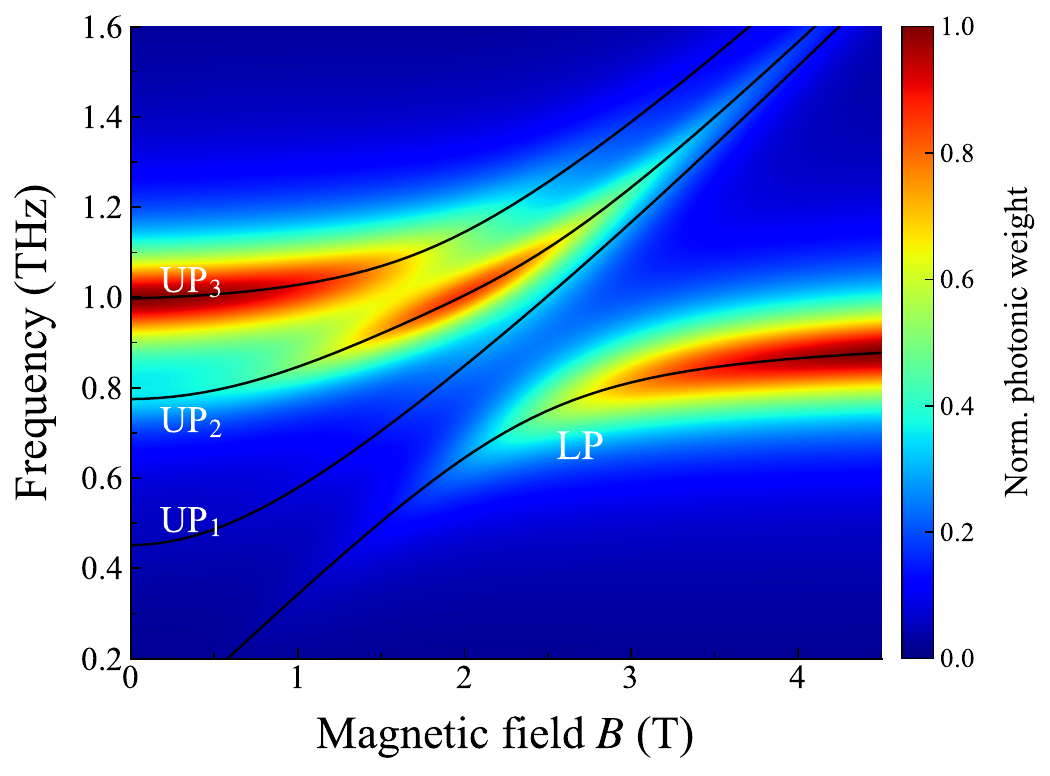}
\caption{
Lorentzian broadened photonic weight map constructed using Eq.~\eqref{eq:synthetic_map}. The map is a phenomenological visibility proxy for the bright polariton branches and is not a full input--output transmission simulation.
}
\label{fig:supp_transmission_map}
\end{figure}

The positive frequency eigenvalues $\Omega_\nu$ of the active chiral block define the polariton branches LP, UP\(_1\), UP\(_2\), and UP\(_3\). For an active sector eigenvector
\begin{equation}
    \bm{v}_{\nu}
    =
    \left(
    w_{\nu,a_+},
    y_{\nu,a_-},
    w_{\nu,\mathrm{CR}},
    w_{\nu,\mathrm{MP1}},
    w_{\nu,\mathrm{MP3}}
    \right)^T,
\end{equation}
the $a_+$ photonic weight by branch is evaluated as
\begin{equation}
    V_{\nu}
    =
    |w_{\nu,a_+}|^2
    -
    |y_{\nu,a_+}|^2=|w_{\nu,a_+}|^2.
    \label{eq:photonic_weight}
\end{equation}
Fig~\ref{fig:supp_photonic_weights} shows $a_+$ net photonic weights \(V_\nu\), Eq.~\eqref{eq:photonic_weight}, by branch in the active spectroscopic sector.
To obtain the transmission-like map, as shown in Fig.~\ref{fig:supp_transmission_map}, each branch is broadened by a Lorentzian and weighted by its photonic content:
\begin{equation}
    I(\omega,B)
    =
    \sum_{\nu\in\{\mathrm{LP},\mathrm{UP}_1,\mathrm{UP}_2,\mathrm{UP}_3\}}
    V_{\nu}(B)
    \frac{\Gamma_\nu}
    {[\omega-\Omega_\nu(B)]^2+\Gamma_\nu^2}.
    \label{eq:synthetic_map}
\end{equation}
The final map is normalized by its maximum value,
\begin{equation}
    \tilde I(\omega,B)=
    \frac{I(\omega,B)}{\max_{\omega,B} I(\omega,B)}.
\end{equation}
The phenomenological broadening parameters used in the plots are
\begin{align}
    \Gamma_{\mathrm{LP}}&=0.10~\mathrm{THz},\quad
    \Gamma_{\mathrm{UP}_1}=0.085~\mathrm{THz},\\
    \Gamma_{\mathrm{UP}_2}&=0.065~\mathrm{THz},\quad
    \Gamma_{\mathrm{UP}_3}=0.10~\mathrm{THz}.
\end{align}
These parameters are chosen only to provide a visually smooth spectroscopic benchmark of the active sector branch structure.

\suppnote{Numerical verification of the chiral selection rule}{supp4}

Here we verify the selection rule for anomalous correlators directly at the level of the ground state covariance matrix. The conserved chiral charge predicts that anomalous correlators with nonzero chiral charge vanish, while charge neutral anomalous channels may remain finite.

The conserved chiral charge is
\begin{equation}
    Q
    =
    a_{+}^{\dagger}a_{+}
    +
    \sum_{j}m_j^{\dagger}m_j
    -
    a_{-}^{\dagger}a_{-},
\end{equation}
where \(m_j\in\{\mathrm{CR},\mathrm{MP}_1,\mathrm{MP}_3\}\). Operators with nonzero \(Q\) charge have vanishing expectation values in a \(Q\)-eigenstate. Therefore,
\begin{equation}
    \langle a_+a_+\rangle=0,
    \qquad
    \langle a_+m_j\rangle=0,
\end{equation}
whereas the \(a_-\)-matter anomalous channels are allowed by symmetry, except for $\langle a_- a_-\rangle$.

\begin{figure}[t]
\centering
\includegraphics[width=\linewidth]{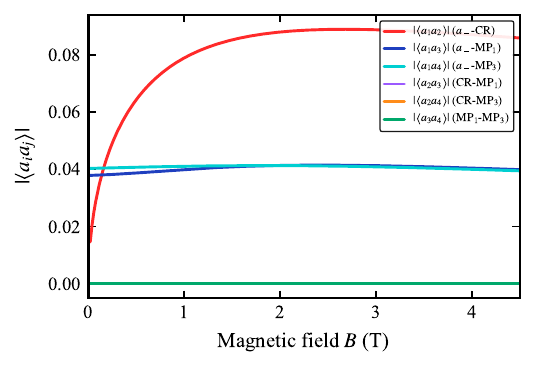}
\caption{
Allowed anomalous correlators as a function of magnetic field. The dominant finite channels involve \(a_-\) and the matter modes, together with the allowed circular cavity anomalous channel.
}
\label{fig:supp_anomalous_norm}
\end{figure}

The anomalous correlation matrix is obtained from the Bogoliubov coefficients as
\begin{equation}
    \langle a_i a_j\rangle =-\sum_\nu w_{\nu,i}^* y_{\nu,j}.
\end{equation}
The numerical results shown in Figure~\ref{fig:supp_anomalous_norm} confirm that all forbidden anomalous correlators remain at the numerical floor, while \(a_-\)-matter channels, allowed by symmetry, are finite.

\suppnote{Quantum correlation measures}{supp5}
\begin{figure*}[t]
\centering
\subfloat[]{%
\includegraphics[width=0.33\linewidth]{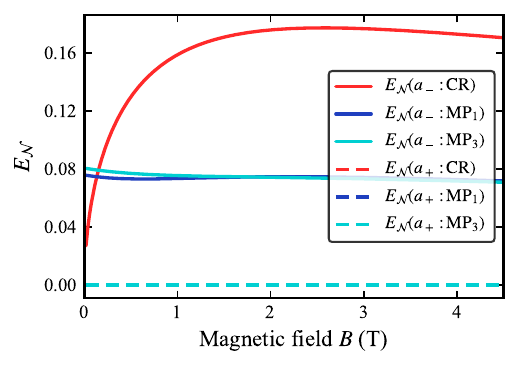}
\label{fig:supp_photon_matter_EN}
}%
\subfloat[]{%
\includegraphics[width=0.33\linewidth]{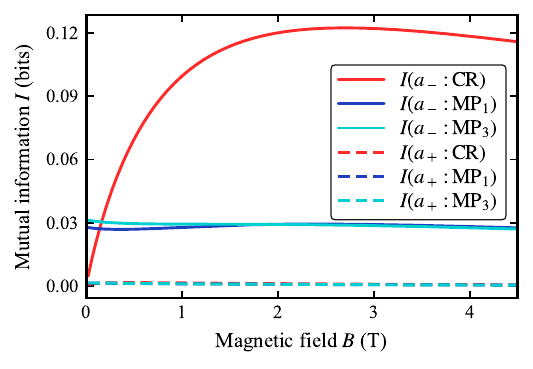}
\label{fig:supp_MI}
}%
\subfloat[]{%
\includegraphics[width=0.33\linewidth]{Discord_photon_matter_vs_B_bits.pdf}
\label{fig:supp_photon_matter_discord}
}

\caption{
Correlations between cavity polarizations and matter modes in the ground state.
(a) Logarithmic negativity, Eq.~\eqref{log_negativity};
(b) mutual information, Eq.~\eqref{eq:mutualinfo};
and (c) Gaussian discord, Eq.~\eqref{gaussian_discord}.
Solid curves correspond to correlations involving \(a_-\), while dashed curves correspond to correlations involving \(a_+\).
Across all three measures, the \(a_-\) polarization carries the dominant cavity--matter correlation content, whereas correlations involving \(a_+\) remain negligible or at the numerical floor.
These results demonstrate a strong polarization selectivity of the correlated ground state and identify \(a_-\) as the cavity channel participating in the light--matter hybridization.
}
\label{fig:supp_cavity_matter_correlations}
\end{figure*}

\begin{figure*}[t]
\centering
\subfloat[]{%
\includegraphics[width=0.33\linewidth]{Discord_matter_pairs_vs_B_bits.pdf}
\label{fig:supp_matter_discord}
}%
\subfloat[]{%
\includegraphics[width=0.33\linewidth]{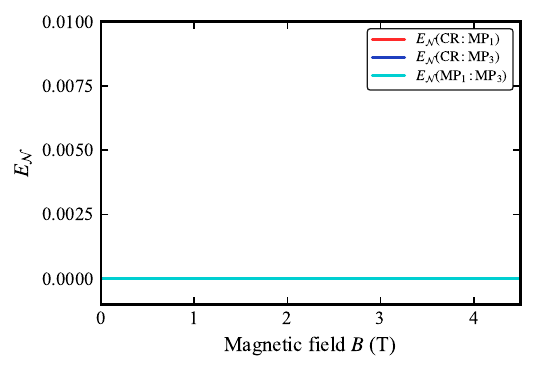}
\label{fig:supp_matter_pairwise_EN}
}%
\subfloat[]{%
\includegraphics[width=0.33\linewidth]{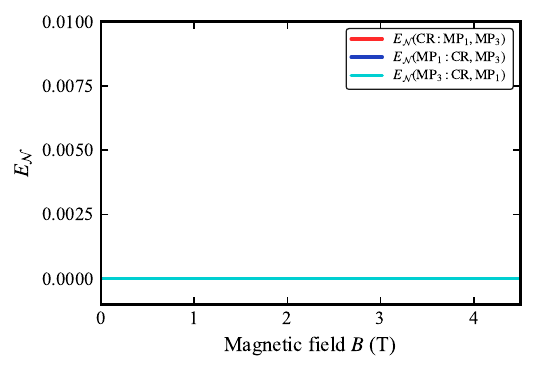}
\label{fig:supp_matter_1vs2_EN}
}

\caption{
Matter--matter correlations within the CR, MP\(_1\), and MP\(_3\) sector.
(a) Gaussian discord between matter-mode pairs.
(b) Pairwise logarithmic negativities for the \(1|1\) cuts CR:MP\(_1\), CR:MP\(_3\), and MP\(_1\):MP\(_3\).
(c) Matter-only \(1|2\) logarithmic negativities for the cuts CR:\(\{\mathrm{MP}_1,\mathrm{MP}_3\}\), MP\(_1\):\(\{\mathrm{CR},\mathrm{MP}_3\}\), and MP\(_3\):\(\{\mathrm{CR},\mathrm{MP}_1\}\).
The finite discord in panel (a), together with the vanishing logarithmic negativities in panels (b) and (c), shows that the matter subsystem retains nonclassical Gaussian correlations without supporting either pairwise or matter-only multipartite entanglement.
}
\label{fig:supp_matter_correlations}
\end{figure*}
We introduce the three principal measures used throughout our analysis to characterize quantum correlations and information in the two mode ground state: logarithmic negativity, mutual information, and quantum discord. Together, these measures demonstrate that the spectroscopically active polarization mode $a_+$ remains effectively decoupled from the matter sector at the level of the ground state, while the dominant vacuum correlations are carried by the $a_-$ channels.

To quantify bipartite entanglement, we employ the logarithmic negativity, which is based on the positive partial transpose (PPT) criterion~\cite{SimonPRL2000,DuanPRL2000}. The logarithmic negativity across $A|B$ partition is computed from the symplectic eigenvalues \(\tilde{\nu}_k\) of the partially transposed covariance matrix restricted to the bipartition,
\begin{equation}
    E_{\mathcal{N}}(A:B)
    =
    \sum_k
    \max\{0,-\log(2\tilde{\nu}_k)\}.
    \label{log_negativity}
\end{equation}
In continuous variable systems, the partial transpose corresponds to flipping the sign of the momentum $(x,p)\rightarrow(x,-p)$ quadratures of subsystem $B$.  

The quantum mutual information quantifies the total correlations between two subsystems:
\begin{equation}
I(A{:}B)=S(\rho_A)+S(\rho_B)-S(\rho_{AB}),
\label{eq:mutualinfo}
\end{equation}
where $S(\rho)$ is the von Neumann entropy defined as:
\begin{equation}
    S(\rho)=-\mathrm{Tr}(\rho \log_2(\rho))
\end{equation}
In the multimode Hopfield system the subsystems $A$ and $B$ correspond to subsets of cavity or matter modes.  
Because the states are Gaussian, each entropy entering $I(A{:}B)$ is obtained from the symplectic eigenvalues $\nu_k$ of the corresponding reduced covariance matrix~\cite{WeedbrookRMP2012}, such as:
\begin{equation}
S(\rho)=\sum_{k} h(\nu_k),
\end{equation}
where the function $h(\nu)$ is given by:
\begin{equation}
h(\nu)=\Big(\nu+\tfrac12\Big)\log_2\Big(\nu+\tfrac12\Big)
-\Big(\nu-\tfrac12\Big)\log_2\Big(\nu-\tfrac12\Big).
\end{equation}
Since the quantum mutual information $I$ captures total correlations, we will use quantum discord $D$ to measure quantum correlations shared by the two subsystems $A$ and $B$~\cite{PhysRevLett.88.017901}. For pure states discord coincides with entanglement, but for mixed states captures quantum correlations shared even when the state is separable \textit{i.e.} no entanglement. 
The quantum discord is defined as the mismatch between two quantities that are equivalent in the classical information theory:
\begin{equation}
    D_{A\leftarrow B}=I(A{:}B)-J_{A\leftarrow B},
\end{equation}
where,
\begin{equation}
J_{A\leftarrow B}=S(\rho_A)-\inf_{\mathcal M_B\in\mathcal G} S(\rho_{A|\mathcal M_B}),
\end{equation}
$\rho_{A|\mathcal M_B}$ is the conditional state of $A$ when we measure $B$ using a measurement $\mathcal M_B$ and $\mathcal G$ denotes the set of measurements on $B$ which will be restricted to Gaussian ones, thus we will refer to it as Gaussian discord~\cite{GiordaParis2010,EisertPlenio2003}.
Unlike the logarithmic negativity or the mutual information, the discord is not symmetric as it depends on the system upon which the measurement acts, therefore, we use the symmetrized Gaussian discord,
\begin{equation}
    D(A:B)
    =
    \frac{1}{2}
    \left[
    D_{A\leftarrow B}
    +
    D_{B\leftarrow A}
    \right].
    \label{gaussian_discord}
\end{equation}
Figures~\ref{fig:supp_photon_matter_EN}, ~\ref{fig:supp_MI} and~\ref{fig:supp_photon_matter_discord} demonstrate that $a_+$ is decoupled from the structure of the ground state, while $a_-$ exhibiting finite correlations with all matter modes and through all measures.




\begin{figure*}[t]
\centering
\subfloat[]{%
\includegraphics[width=0.33\linewidth]{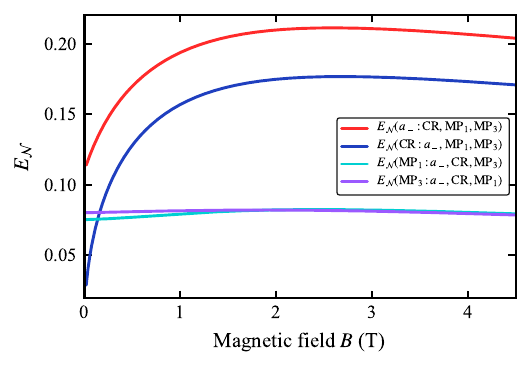}
\label{fig:supp_1vs3}
}%
\subfloat[]{%
\includegraphics[width=0.33\linewidth]{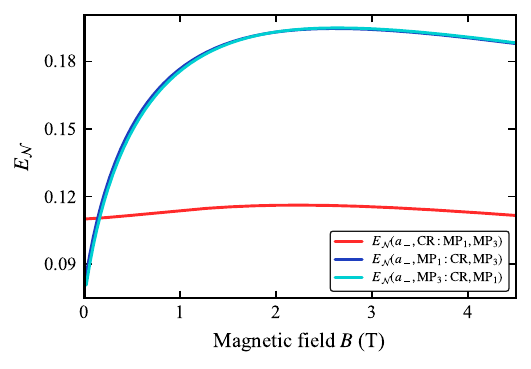}
\label{fig:supp_2vs2}
}%
\subfloat[]{%
\includegraphics[width=0.33\linewidth]{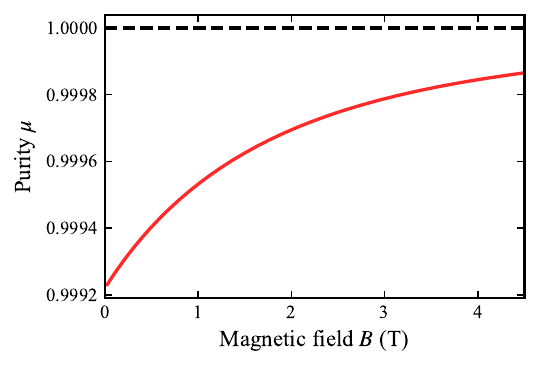}
\label{fig:supp_purity}
}
\caption{
Entanglement structure and purity of the reduced four mode subsystem
\(\{a_-,\mathrm{CR},\mathrm{MP}_1,\mathrm{MP}_3\}\).
(a) Logarithmic negativity, Eq.~\eqref{log_negativity}, across all \(1|3\) bipartitions.
(b) Logarithmic negativity, Eq.~\eqref{log_negativity}, across all \(2|2\) bipartitions.
(c) Purity, Eq.~\eqref{purity}, of the reduced four mode state.
The finite logarithmic negativities observed for both classes of bipartitions demonstrate that the cavity mode \(a_-\) is genuinely entangled with the collective matter sector across multiple partitions. Simultaneously, the purity remains close to unity over the entire parameter range, indicating that tracing out the inactive polarization \(a_+\) introduces only negligible mixedness. Together, these results show that the essential quantum-correlated ground-state structure is faithfully captured by the nearly pure four mode subsystem, validating its use as an effective description of the light--matter entanglement.
}
\label{fig:supp_4mode_entanglement}
\end{figure*}

As a result, the vanishing \(a_+\)-matter entanglement and the negligible \(a_+\)-matter mutual information and discord show that \(a_+\) is effectively decoupled from the matter sector in the ground state correlation structure. This explains why tracing out \(a_+\) introduces very little mixedness into the reduced \(\{a_-,\mathrm{CR},\mathrm{MP}_1,\mathrm{MP}_3\}\) sector. Furthermore, we establish that the matter subsystem alone does not contain pairwise or tripartite entanglement, even though it carries finite Gaussian discord as shown in Fig.~\ref{fig:supp_matter_discord}. This distinction is central to the interpretation of the cavity as mediating nonclassical but non-entangling matter correlations. For the three matter modes \(\{\mathrm{CR},\mathrm{MP}_1,\mathrm{MP}_3\}\), we test all pairwise \(1|1\) cuts (Fig.~\ref{fig:supp_matter_pairwise_EN}) and all \(1|2\) bipartitions (Fig.~\ref{fig:supp_matter_1vs2_EN}) using logarithmic negativity. The pairwise checks rule out direct two-mode matter entanglement. The \(1|2\) cuts rule out tripartite entanglement within the matter subsystem. Thus, within the matter subsystem, the cavity-mediated correlations appear as Gaussian discord rather than entanglement. This supports the interpretation in the main text that the relevant correlated object is not a matter-only subsystem, but the collective \(a_-\)-centered sector.




\suppnote{Multipartite entanglement, residual photonic correlations, and purity}{supp7}

Here we analyze the reduced four mode subsystem
\begin{equation}
    \mathcal{S}_4
    =
    \{a_-,\mathrm{CR},\mathrm{MP}_1,\mathrm{MP}_3\}.
\end{equation}
The goal is to show that the dominant correlated sector is nearly isolated, effectively multipartite entangled, and only weakly mixed by the residual coupling to \(a_+\). Fig.~\ref{fig:supp_1vs3} and Fig.~\ref{fig:supp_2vs2} show, respectively, that all inequivalent \(1|3\) \(2|2\) bipartitions exhibit robust inseparability, which  that the reduced state of subsystem $\mathcal{S}_4$ fully inseparable.



\begin{figure}[b]
\centering
\includegraphics[width=\linewidth]{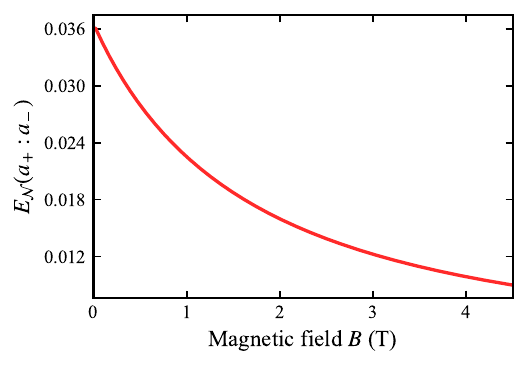}
\caption{
Residual logarithmic negativity between the two cavity polarizations \(a_+\) and \(a_-\). The entanglement is small compared with the dominant \(a_-\)-matter channels and decreases with magnetic field.
}
\label{fig:supp_photon_photon_EN}
\end{figure}

The purity of a Gaussian state with covariance matrix \(V\) is given by~\cite{WeedbrookRMP2012}:
\begin{equation}
    \mu
    =
    \frac{1}{\sqrt{\det(2V)}}.
    \label{purity}
\end{equation}
For the reduced four mode state, the purity remains close to unity over the magnetic field range as shown in Fig.~\ref{fig:supp_purity}, supporting the interpretation of the correlated subsystem as an approximately isolated Gaussian sector. Moreover, since the state of $\mathcal{S}_4$ is fully inseparable and taking its high purity, we can claim that the reduced four mode state effectively exhibits genuine multipartite entanglement~\cite{GUHNE20091}.
The residual entanglement between the two circular cavity modes is small and decreases with magnetic field. This further supports the approximate factorization between the weakly connected active polarization and the dominant \(a_-\)-centered correlated sector. The bipartition negativities, near-unity purity, and weak residual \(a_+\)-\(a_-\) entanglement (as shown in Fig.~\ref{fig:supp_photon_photon_EN}) show that \(\mathcal{S}_4\) behaves as a nearly isolated \(a_-\)-centered correlated sector of the full ground state.

\begin{figure*}
\centering
\subfloat[]{%
\includegraphics[width=0.33\linewidth]{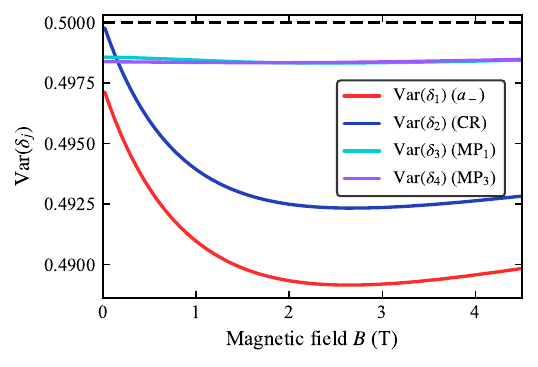}
\label{fig:supp_nullifier_variances}
}%
\subfloat[]{%
\includegraphics[width=0.33\linewidth]{nullifier_squeezing_db_vs_B.pdf}
\label{fig:supp_nullifier_squeezing}
}%
\subfloat[]{%
\includegraphics[width=0.33\linewidth]{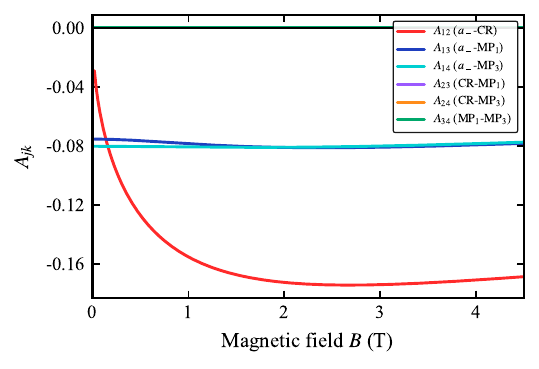}
\label{fig:supp_adjacency}
}

\caption{
Continuous variable graph-state diagnostics of the reduced four mode subsystem
\(\{a_-,\mathrm{CR},\mathrm{MP}_1,\mathrm{MP}_3\}\).
(a) Optimized nullifier variances, with the dashed line indicating the vacuum threshold \(1/2\).
(b) Corresponding nullifier squeezing, defined as
\(-10\log_{10}(2\,\mathrm{Var}\,\delta_j)\).
(c) Off-diagonal elements of the effective adjacency matrix.
The suppression of all optimized nullifier variances below the vacuum limit, together with the resulting positive squeezing, demonstrates the emergence of multipartite continuous-variable correlations characteristic of a Gaussian graph state. The extracted adjacency matrix reveals that these correlations are organized in a predominantly star-like topology, with the cavity mode \(a_-\) acting as the central hub and the strongest connection occurring along the \(a_-\)--CR link. These results identify the correlated ground state as a highly structured multimode Gaussian resource whose entanglement architecture is mediated by the active cavity polarization.
}
\label{fig:supp_graph_state_properties}
\end{figure*}

\suppnote{Gaussian graph-state diagnostics}{supp8}

This note explains how the effective graph-state diagnostics used in the main text are obtained from the reduced covariance matrix. The purpose is not to claim an ideal infinitely squeezed cluster state, but to show that the nearly pure correlated sector has an organized Gaussian graph-like structure.

For the reduced subsystem \(\mathcal{S}_4\), an effective continuous variable graph is obtained by minimizing the nullifier variances~\cite{MenicucciPRA2011}
\begin{equation}
    \delta_j
    =
    p_j
    -
    \sum_k A_{jk}q_k.
    \label{nullifier}
\end{equation}
For a covariance matrix written in block form
\begin{equation}
    V
    =
    \begin{pmatrix}
    V_{qq} & V_{qp}\\
    V_{pq} & V_{pp}
    \end{pmatrix},
\end{equation}
the least-squares adjacency matrix is
\begin{equation}
    A_{\mathrm{raw}}
    =
    V_{pq}V_{qq}^{-1}.
\end{equation}
In the numerical implementation, the final reported adjacency matrix is symmetrized as
\begin{equation}
    A
    =
    \frac{1}{2}
    \left(
    A_{\mathrm{raw}}+A_{\mathrm{raw}}^{T}
    \right).
\end{equation}
The nullifier covariance matrix is
\begin{equation}
    V_{\delta}
    =
    V_{pp}
    -
    A V_{qp}
    -
    V_{pq}A^{T}
    +
    A V_{qq}A^{T}.
\end{equation}
The nullifier variances are the diagonal entries of \(V_{\delta}\).

Figure~\ref{fig:supp_graph_state_properties}\subref{fig:supp_nullifier_variances} shows the optimized nullifier variances of the reduced four mode subsystem. All variances remain below the vacuum threshold (1/2), indicating the presence of nontrivial collective quantum correlations among the modes. The same behavior is reflected in the corresponding nullifier squeezing shown in Fig.~\ref{fig:supp_graph_state_properties}\subref{fig:supp_nullifier_squeezing}, where positive squeezing is obtained throughout the parameter range. Finally, Fig.~\ref{fig:supp_graph_state_properties}\subref{fig:supp_adjacency} displays the off-diagonal elements of the effective adjacency matrix. The dominant couplings connect the cavity mode ($a_-$) to the matter modes, with the strongest interaction occurring between ($a_-$) and CR. This structure indicates that the cavity mode acts as the main mediator of correlations within the reduced subsystem and gives rise to a predominantly star-like connectivity pattern.





\end{document}